\documentclass[aip,reprint]{revtex4-1}
\usepackage{epsfig,amssymb,amsmath}
\usepackage{xcolor}
\draft % marks overfull lines with a black rule on the right

\begin{document}
	
	%\preprint{}
	
	\title{Analytical criteria for magnetization reversal in $\varphi_0$ Josephson junction} %Title of paper
	
	

	%~~~~~~~~~~~~~~~~~~~~~~~~~~~~~~~~
	%\author{Yu. M. Shukrinov~$^{1,2}$}
	%\author{I. R. Rahmonov~$^{1,3}$}
	%\author{K. Sengupta$^{4}$ }
	%\author{M. Mikhailova~$^{3}$}
	%\author{Ja. Martincova~$^{5}$}
	%\author{A. Buzdin$^{6}$}
	%~~~~~~~~~~~~~~~~~~~~~~~~~~~~~~~~~~~~
	
	\author{A.~A.~Mazanik}
	\email[]{mazanik@theor.jinr.ru}
	%\thanks{}
	\affiliation{BLTP, JINR, Dubna, 141980, Moscow Region, Russia}
	\affiliation{MIPT, Dolgoprudny, 141700, Moscow Region, Russia}
	
	\author{I.~R.~Rahmonov}
	\email[]{rahmonov@theor.jinr.ru}
	%\thanks{}
	\affiliation{BLTP, JINR, Dubna, 141980, Moscow Region, Russia}
	\affiliation{Umarov Physical Technical Institute, TAS, Dushanbe, 734063, Tajikistan}

\author{A.~E.~Botha}
	\email[]{Bothaae@unisa.ac.za}
	%\thanks{}
	\affiliation{Department of Physics, University of South Africa, Florida 1710, South Africa}
	
	\author{Yu.~M.~Shukrinov}
	\email[]{shukrinv@theor.jinr.ru}
	%\thanks{}
	\affiliation{BLTP, JINR, Dubna, 141980, Moscow Region, Russia}
    \affiliation{Department of Physics, University of South Africa, Florida 1710, South Africa}
    	\affiliation{Dubna State University, Dubna, Moscow Region, 141980, Russia}

	\date{\today}
	
	\begin{abstract}
		The $\varphi_0$ Josephson junctions formed by ordinary superconductors and a magnetic non-centrosymmetric interlayer are studied. We derive an analytical solution for the magnetization dynamics induced by an arbitrary current pulse  and formulate the criteria for magnetization reversal. Using the obtained results, the form and duration of the current pulse are optimized. The agreement between analytical and numerical investigations is reached in the case of a large  product of the ratio Josephson to magnetic energy, strength of spin-orbit interaction and a minimal value of the current pulse. The obtained results allow one to predict magnetization reversal at the chosen system parameters.
	\end{abstract}
	
	\maketitle
	\date{\today}
	
	\section{Introduction}
The ability to manipulate  magnetic properties by the Josephson current and its
	opposite, {\it i.e.}, to influence the Josephson current by magnetic
	moment, has attracted much recent attention.~\cite{linder15,efetov11,buzdin05,bergeret05,golubov04,ghosh17} In the
	superconductor /ferromagnet/ superconductor (SFS) Josephson
	junctions, the spin-orbit interaction in a ferromagnet without
	inversion symmetry provides a mechanism for a direct (linear)
	coupling between the magnetic moment and the superconducting
	current. In such junctions with noncentrosymmetric ferromagnetic interlayer and broken time reversal symmetry, called
	$\varphi_0$ junctions, the current-phase relation (CPR) is given by $I = I_c \sin (\varphi-\varphi_0)$, where the phase
	shift $\varphi_0$ is proportional to the magnetic moment
	perpendicular to the gradient of the asymmetric spin-orbit potential.~\cite{buzdin08}
	
	The $\varphi_0$  junctions lead to the anomalous Josephson effect in different hybrid heterostructures which reflects the play of superconductivity, spin-orbit interactions and magnetism at the same time \cite{yokoyama-prb89,minutillo-prb98,krive-prb71,reynoso-prb101,alidoust-prb96,Alidoust-prb98-085414,	braude-prl98,zyuzin-prb93,zyuzin-prb61,Alidoust-prb98-245418,goldobin-prl107,goldobin-prb91,Menditto-prb98,alidoust-prb87,Shapiro-prb98,spanslatt-prb98} and demonstrate a number of unique features important for superconducting spintronics and modern informational technologies. They  allow one to manipulate
	the internal magnetic moment using the Josephson current.~\cite{buzdin08,konschelle09} Thus, once the magnetization rotates, a reverse phenomenon should be expected. Namely, it might pumps current through the
	$\varphi_0$  phase shift which is fueled by the term proportional to magnetization and spin-orbit coupling. It leads to the appearance of the DC component of superconducting current, playing an important role in the transformation of IV-characteristics in the resonance region.\cite{shukrinov-prb19}

The application of DC voltage to the $\varphi_0$ junction produce current oscillations and consequently magnetic precession.
	As shown in Ref.\cite{konschelle09}, this precession may be monitored by the appearance of higher
	harmonics in the CPR as well as by the presence of a DC component of
	the superconducting current that increases substantially near the ferromagnetic
	resonance (FMR). The authors stressed that the magnetic dynamics of the $SFS$ $\varphi_0$ junction may be quite complicated and strongly anharmonic. In contrast to these results, in Ref. \cite{shukrinov-prb19} has been demonstrated that precession of the magnetic moment  in some current intervals along IV-characteristics may be very simple and harmonic.   It is expected that external
	radiation would lead to a series of novel phenomena. Out of this, the
	possibility of appearance  of half-integer Shapiro steps (in
	addition to the conventional integer steps) and  the generation of
	an additional magnetic precession with frequency of external
	radiation was already discussed in Ref.\ \cite{konschelle09}.
		
An investigation of the heterostructures that combine superconducting  and ferromagnetic  elements gives insight into the problem of the mutual influence of superconductivity and ferromagnetism, allows a realization of exotic superconducting states such as the Larkin-Ovchinnikov-Fulde-Ferrell state and triplet ordering, and promises applications that utilize the spin degree of freedom \cite{braude-prl98}. The possibility of anomalous Josephson effect in SNS junctions can be expected where the normal region is a heterostructure formed by alternating ferromagnetic and spin-orbit coupled segments.\cite{minutillo-prb98} The Josephson junctions composed of two semiconducting nanowires with Rashba spin-orbit coupling and induced superconductivity from the proximity effect display a geometrically induced anomalous Josephson effect, the flow of a supercurrent in the absence of external phase bias.\cite{spanslatt-prb98} A generic nonaligned Josephson junction in the presence of an external magnetic field reveals an unusual flux-dependent current-phase relation.\cite{alidoust-prb87} Such nonaligned Josephson junctions can be utilized to obtain a ground state other than $0$ and $\pi$, corresponding to the $\varphi$ junction, which is tunable via the external magnetic flux. A tunable $\pm \varphi$ and hybrid system between $\varphi$ and $\varphi_0$ junctions were investigated in Refs. \cite{Menditto-prb98,goldobin-prb91,goldobin-prl107}.
		
	Recently, an anomalous phase shift was experimentally observed in different systems, particularly, in the $\varphi_0$  junction based on a nanowire quantum dot \cite{szombati16}.  A quantum interferometer device was used in order to investigate phase offsets and demonstrate that $\varphi_0$  can be controlled by electrostatic gating.  The presence of an anomalous phase shift of $\varphi_0$  was also experimentally observed directly through CPR measurement in a hybrid SNS JJ fabricated using ${\rm Bi_2Se_3}$ (which is a topological insulator with strong spin-orbit coupling) in the presence of an in-plane magnetic field~\cite{aprili19}. This constitutes a direct experimental measurement of the spin-orbit coupling strength and opens up new possibilities for phase-controlled Josephson devices made from
	materials with strong spin-orbit coupling. In Ref.\cite{chudn2016,chudn2010}, the authors argued that the $\varphi_0$  Josephson junction is ideally suited for studying of quantum tunneling of the magnetic moment. They proposed that magnetic tunneling would show up in the ac voltage across the junction and it could be controlled by the bias current applied to the junction.  Though the static properties of the SFS structures are well studied both theoretically and experimentally, much less is known about the magnetic dynamics of these systems~\cite{waintal02,braude08,linder83}. The observation of a tunable anomalous Josephson effect in InAs/Al Josephson junctions measured via a superconducting quantum interference device (SQUID) reported in Ref.\cite{mayer19} By gate controlling the density of InAs  the authors were able to tune the spin-orbit coupling of the Josephson junction by more than one order of magnitude. This gives  the ability to tune $\varphi_0$, and opens several new opportunities for superconducting spintronics \cite{linder15}, and new possibilities for realizing and characterizing topological superconductivity \cite{alicea12,fornieri19,ren19}.

	One of the milestones for superconducting electronics which stands out by ultra-low  energy dissipation is the creation of cryogenic memory \cite{herr11,baek14,mukhanov11}. Different realizations for such devices were proposed including devices based on the $\varphi_0$ Josephson junctions \cite{baek14,birge15,nguyen2019,bergeret19}.  The DC superconducting current applied to a SFS $\varphi_0$ junction might produce a strong orientation effect on the ferromagnetic layered magnetic moment~\cite{apl17}.  The full magnetization reversal can be realized  by applying an electric current pulse.\cite{apl17} Detailed pictures representing the intervals of the damping parameter $\alpha$, Josephson to magnetic energy relation $G$ and the spin-orbit coupling parameter $r$ were obtained with the full magnetization reversal~\cite{jetpl-atanas19}. It was demonstrated that the appearance of the  reversal was sensitive to changing the system parameters and showed some periodic structure. Guarcello and Bergeret in Ref.\ \onlinecite{bergeret19} suggested to use a $\varphi_0$ SFS junction as a cryogenic memory based on the current pulse switching of magnetization proposed in Ref.\ \onlinecite{apl17}. In this scheme a bit of information is associated with the direction of the magnetic moment along or opposite the direction of the easy axis of the ferromagnetic layer. The writing is carried out as a reversal of the magnetic moment by a pulse of current and the readout is performed by detection of the magnetic flux by SQUID inductively coupled to the $\varphi_0$ junction. They also explored the robustness of the current-induced magnetization reversal against thermal fluctuations and suggested a way of decoupling the Josephson phase and the magnetization dynamics by tuning the Rashba spin-orbit interaction strength via a gate voltage. A suitable non-destructive readout scheme based on dc-SQUID inductively coupled to the $\varphi_0$ junction was discussed as well. We stress that  in all the above mentioned works the magnetization reversal  was studied numerically only.
	
	In this work we derive an analytical solution for the magnetization dynamics induced by an arbitrary current pulse  and formulate the criteria for magnetization reversal in the $\varphi_0$ Josephson junctions formed by ordinary superconductors and a magnetic, non-centrosymmetric interlayer. Using the obtained analytical results, we optimize the form and duration of the current pulse. The agreement between analytical and numerical investigations is reached in the case of a large  product of the ratio of the Josephson energy to the magnetic energy, strength of spin-orbit interaction and a minimum value of the flowing current. The obtained results explains  the periodicity in the appearance of the magnetization reversal intervals observed in Ref.\ \onlinecite{jetpl-atanas19} and allow one to predict magnetization reversal at the chosen system parameters.
	
The plan of the rest of this work is as follows. In Sec. II	we introduce the model and methods, particularly, the derivation of effective field and current pulse. This is followed by Sec. III, where the relation between expressions for the temporal dependence of the current's pulse  and the superconducting current is obtained for different ratio of Josephson to characteristic frequency of junction. In  Sec. IV we present solution of Landau-Lifshiz-Gilbert equation for the case, when product of the ration of Josephson to magnetic energy and spin-orbit coupling is much more that one. Section V is devoted to small damping regime. We discuss the periodicity in the appearance of the magnetization reversal intervals in the diagrams ``Gilbert damping - ration of Josephson to magnetic energy''. The periodicity for the diagram ``spin-orbit coupling - ration of Josephson to magnetic energy'' is discussed in Sec.VI. Finally, in Sec.VII we summarize our main results and conclude.
	
	\section{Model and Methods}
	
Physics of SFS Josephson structures is determined by system of equations which consist of Landau-Lifshits-Gilbert (LLG), resistively shunted junction (RSJ) model,  and Josephson relation between phase difference and voltage. The dynamics of the magnetic moment in $\varphi_0$ JJ is described by the LLG equation \cite{lifshitz91}
	\begin{equation}\label{eq:LLG}
	\frac{d \mathbf{M}}{dt} = \gamma_m \mathbf{H}_{eff}\times\mathbf{M} + \frac{\alpha}{M_0} \left(\mathbf{M}\times \frac{d \mathbf{M}}{dt}\right),
	\end{equation}
	\noindent where $\mathbf{M}$ is the magnetization vector, $\gamma_m$ is the gyromagnetic relation, $\mathbf{H}_{eff}$ is the effective magnetic field, $\alpha$ is Gilbert damping parameter, $M_0=|\mathbf{M}|$.
	
	In order to find the expression for the effective magnetic field we have used the model developed in Ref.\ \onlinecite{konschelle09}, where it is assumed that the gradient of the spin-orbit potential is along the easy axis of magnetization taken to be along ${\hat z}$. In this case the total energy of the
	system can be written as
	\begin{equation}
	E_{\text{tot}}=-\frac{\Phi_{0}}{2\pi}\varphi I+E_{s}\left(  \varphi,\varphi_{0}\right)  +E_{M}\left(  \varphi_{0}\right)  , \label{EQ_energy}%
	\end{equation}
	where $\varphi$ is the phase difference  between the
	superconductors across the junction, $I$ is the external current,
	$E_{s}\left( \varphi,\varphi_{0}\right)=E_{J}\left[  1-\cos\left(
	\varphi-\varphi_{0}\right)  \right]$, and $\displaystyle
	E_{J}=\Phi_{0}I_{c}/2\pi$ is the Josephson energy. Here $\Phi_{0}$
	is the flux quantum, $I_{c}$ is the critical current, $\varphi_{0}=l
	\upsilon_{so} M_y/(\upsilon_{F} M_{0})$, $l=4 h L/\hbar
	\upsilon_{F}$, $L$ is the length of $F$ layer, $h$ is the exchange
	field of the $F$ layer, $E_{M}=-K\mathcal{V}M^2_z/(2 M_0^2)$, the
	parameter $\upsilon_{so}/\upsilon_{F}$ characterizes a relative
	strength of spin-orbit interaction, $K$ is the anisotropic constant,
	and $\mathcal{V}$ is the volume of the ferromagnetic ($F$) layer.
	
	Consequently the effective field is determined by
	\begin{eqnarray} \label{eq:Heff}
	{\bf H_{eff}}&=&-\frac{1}{\mathcal V}\frac{\partial E_{tot}}{\partial
		{\bf M}}\nonumber\\
	&=&\frac{K}{M_{0}}\bigg[G r \sin\bigg(\varphi - r
	\frac{M_{y}}{M_{0}} \bigg) {\bf\widehat{y}} +
	\frac{M_{z}}{M_{0}}{\bf\widehat{z}}\bigg]
	\label{effective_field}
	\end{eqnarray}
	where $r=l\upsilon_{so}/\upsilon_{F}$, and $\displaystyle G=
	E_{J}/(K \mathcal{V})$.
	
	Using (\ref{eq:LLG}) and (\ref{effective_field}),
	we obtain the system of equations, which describes the dynamics of the magnetization of F layer in SFS structure
\begin{equation}
\label{syseq}
\begin{array}{llll}
\displaystyle \dot{m}_{x}=\frac{1}{1+\alpha^{2}}\{-m_{y}m_{z}+Grm_{z}\sin(\varphi -rm_{y})\\
-\alpha[m_{x}m_{z}^{2}+Grm_{x}m_{y}\sin(\varphi -rm_{y})]\},
\vspace{0.2 cm}\\
\displaystyle \dot{m}_{y}=\frac{1}{1+\alpha^{2}}\{m_{x}m_{z}\\
-\alpha[m_{y}m_{z}^{2}-Gr(m_{z}^{2}+m_{x}^{2})\sin(\varphi -rm_{y})]\},
\vspace{0.2 cm}\\
\displaystyle \dot{m}_{z}=\frac{1}{1+\alpha^{2}}\{-Grm_{x}\sin(\varphi -rm_{y})\\
-\alpha[Grm_{y}m_{z}\sin(\varphi -rm_{y})-m_{z}(m_{x}^{2}+m_{y}^{2})]\},
\end{array}
\end{equation}
where $m_{x,y,z} = M_{x,y,z}/M_0$ satisfy the constraint
	$\sum_{i=x,y,z} m_{i}^2(t)=1$. In this system of equations
	time is normalized to the inverse ferromagnetic resonance frequency
	$\omega_{F}=\gamma K/M_{0}: (t\rightarrow t \omega_F)$.
	
In order to describe the full dynamics of SFS structure the LLG equations should be supplemented by the equation for phase difference $\varphi$, i.e. equation of RSJ model.
According to the extended RSJ model \cite{bobkovi19}, which takes into account derivative of $\varphi_{0}$ phase shift, the current flowing through the system in overdamped case is determined by
\begin{equation}\label{eq:RSJM}
	I =  \frac{\hbar}{2eR}\left[ \frac{d \varphi}{ d t} - \frac{r}{M_{0}} \frac{d M_y}{d t}\right] + I_{c}\sin{\left(\varphi - \frac{r}{M_{0}} M_y\right)}.
	\end{equation}
or in the normalized variables it takes a form
	\begin{equation}\label{eq:rsj}
	I = w  \frac{d \Phi}{ d t}  + \sin{\Phi}.
	\end{equation}
where the bias current $I$ is normalized to the critical one $I_{c}$ and $\Phi = \varphi - r m_y$.

We note that in order to use the same time scale in the LLG and RSJ equations we have normalized time to the $\omega^{-1}_F$. In this case a new parameter determined by the $w=\omega_{F}/\omega_{c}$ is occurred, were $\omega_{c}=\frac{2 e I_c R}{\hbar}$ is a characteristic frequency of the Josephson junction. As we will see, the behaviour of the system depend on the value of this parameter and it characterizes its different regimes.

The current pulse is $I=I_{p}$ and its has rectangular form
	\begin{eqnarray}
	\label{eq:pulse}
	I_{p}(t) = \left\{
	\begin{array}{ll}
	A_s, & t \in [t_0 , t_0 +  \delta t];\\
	0, & \textrm{otherwise},
	\end{array}
	\right.
	\end{eqnarray}
where $A_s$ and $\delta t$ are the pulse amplitude and width, respectively. We note that if any other form is not specified below, the rectangular form is considered.
		
The initial conditions for the LLG equation are the $m_x(t=0)=0$, $m_y(t=0)=0$, $m_z(t=0)=1$ and for RSJ-model equation  $\varphi(t=0)=0$. Via the numerical solution\cite{apl17,jetpl-atanas19} of Eq.(\ref{syseq})  taking into account (\ref{eq:rsj}) and  (\ref{eq:pulse}) we obtain the time dependence of magnetization $m_{x,y,z}(t)$, phase difference $\varphi(t)$ and normalized superconducting current $I_{s}(t) = \sin(\varphi-rm_{y})$
	
In this paper we also compare the analytical and numerical results concerning the periodicity of  magnetization reversal  in $\alpha-G$ and $r-G$ planes.  In order to demonstrate the realization of the magnetization reversal intervals, we solve numerically system of differential equation (\ref{syseq}) for the fixed values of $G$ and $\alpha$ (or $G$ and $r$). Then we have checked the value of $m_{z}$ at the end of each time domain, and if the reversal is realized, the values of $G$ and $\alpha$ (or $r$)  are recorded to the files. Repeating this procedure for the different values of our parameters we build the figures which demonstrates the magnetization reversal appearance in $\alpha-G$ and $r-G$ planes.

	\section{Relation between $I_p(t)$ and $\sin{\Phi(t)}$ at different $w$}
As we mentioned above, the physics of switching in the  $\varphi_0$  is determined by LLG equation (\ref{eq:LLG}) and RSJ model equation (\ref{eq:rsj}). 	An interesting feature of this system of equations in the overdamped case is a decoupling, i.e., equation (\ref{eq:rsj}) for $\Phi$ is decoupled from LLG equation (\ref{syseq}). It allows to find the analytical solution for $\Phi$ and build the theory for magnetization reversal at some values of model and pulse parameters.
	
To investigate the magnetization dynamics, one should start with the solving equation (\ref{eq:rsj}) for the pulse $I_p(t)$ and calculating $\sin\Phi$, which is needed to determine the effective magnetic field (\ref{effective_field}). One will find that the $\sin{\Phi}$ profile consists of two regions: the first one  is the pumping of the $\sin{\Phi}$ during the pulse and the second one is the dropping $\sin{\Phi}$ to zero when the pulse has been switched off.  We investigate these processes in the case of the rectangular pulse (\ref{eq:pulse}). During the pulse $t_0 \leq t \leq t_0 + \delta t $, the equation for $\Phi$ has a form
	\begin{equation}%\label{eq:rsj}
	A_{s} =   w  \frac{d \Phi}{ d t}  + \sin{\Phi},
	\end{equation}
	with the initial  condition $\Phi(t = t_0) = 0$.  For $A_s < 1$ it gives
	\begin{equation}\label{eq:tanp}
	\tan{\Phi(t)/2} =  A_{s}
	\frac{ \tanh{\left[\frac{t-t_0}{\tau_0}\right]}}{\tanh{\left[\frac{t-t_0}{\tau_0}\right]} + \sqrt{1 - A_{s}^{2}} }.
	\end{equation}
	Here $\tau^{-1}_0 = \frac{\sqrt{1 - A_{s}^{2}}}{2w}$,  which determines time scale for approaching  a constant value of $\Phi$. So, the formula (\ref{eq:tanp}) allows to calculate the $\sin{\Phi}$ during the pulse $t_0 \leq t \leq t_0 + \delta t $.
	
	In the second region $t \geq t_0 + \delta t $, when the pulse has been switched off $(I_p(t) = 0)$, we have
	\begin{equation}\label{eq:sin}
	\sin{\Phi(t)} = \frac{2 \tan{\left(\frac{\Phi(t_0 + \delta t)}{2}\right)} }{\exp{\left[\frac{t -t_0 -\delta t}{w}\right]} + \tan^2{\left(\frac{\Phi(t_0 + \delta t)}{2}\right)} \exp{\left[-\frac{t-t_0-\delta t}{w}\right]} }
	\end{equation}
	which exponentially drops to zero with a time scale $\tau_1  \sim w$. Here $\tan\left[\frac{\Phi(t_0+\delta t)}{2}\right]$ is determined by the equation (\ref{eq:tanp}).
	
\begin{figure}[h]
		\centering	\includegraphics[width=0.25\textwidth]{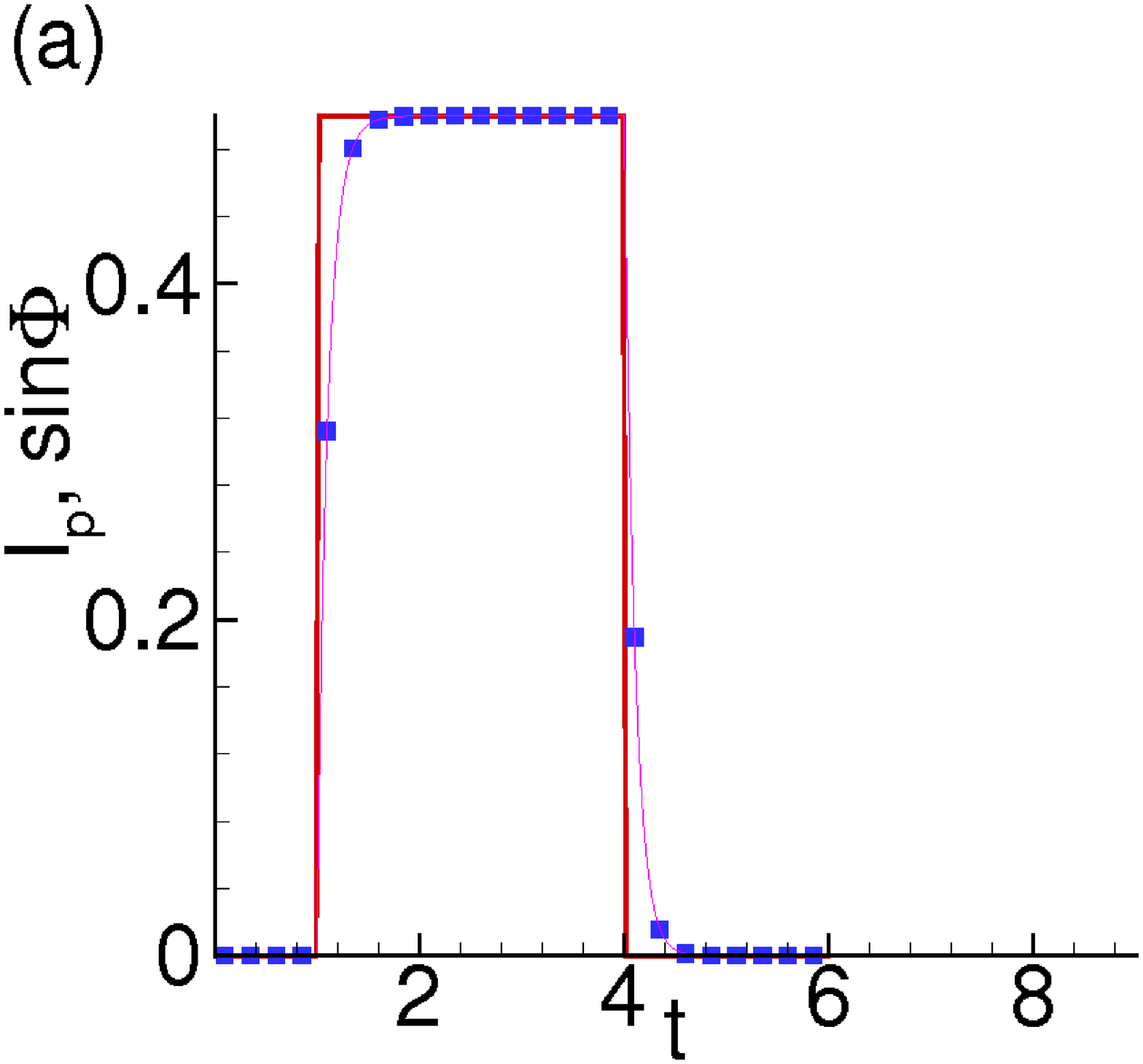}\includegraphics[width=0.25\textwidth]{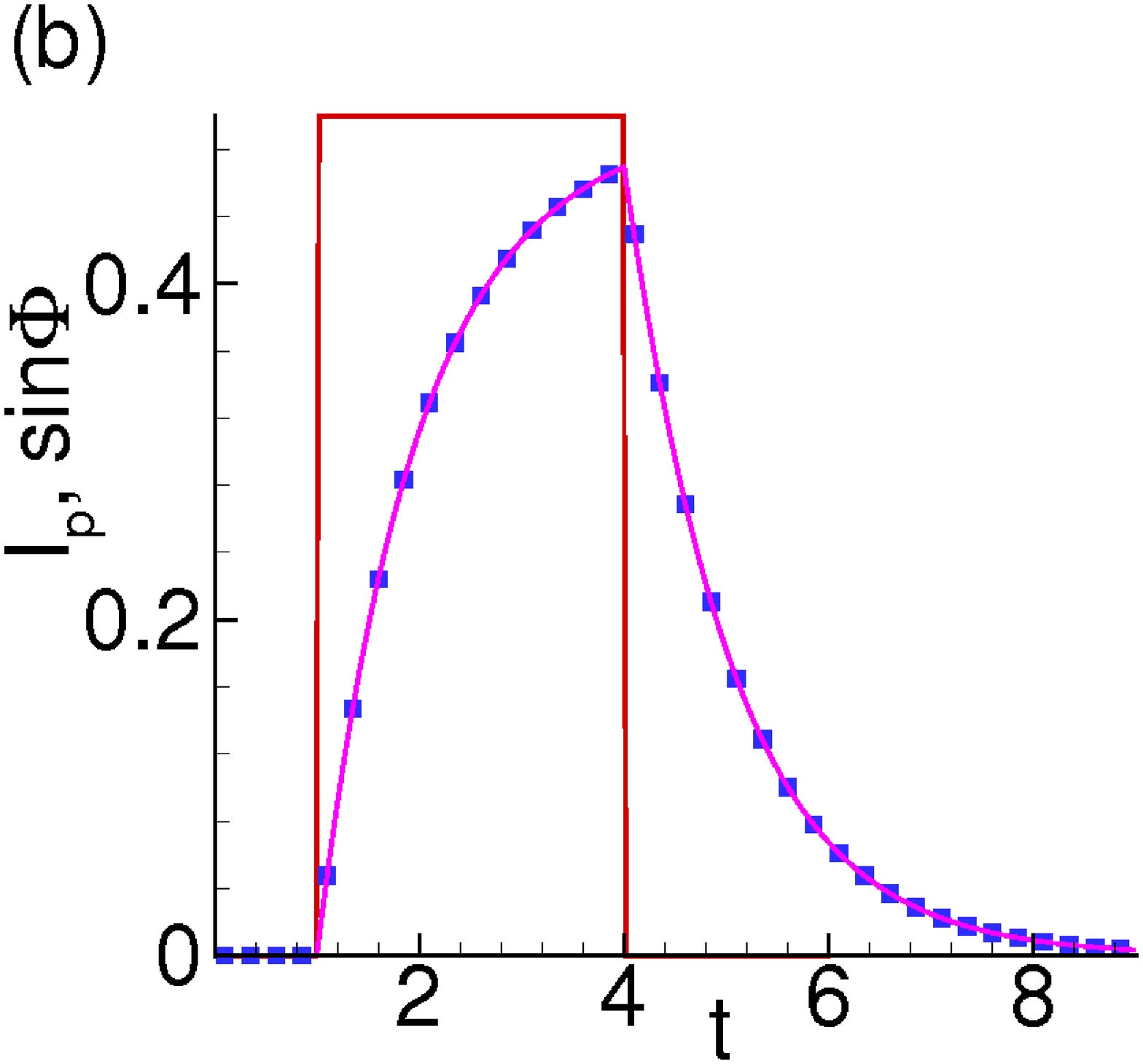}
		\caption{\label{fig:sinPhiIp} (a) Dynamics of $\sin{\Phi}$ (blue, dashed line), based on the formulas (\ref{eq:tanp}) and (\ref{eq:sin}), and   $I_p(t)$ (red line) at $w = 0.1$. The parameters of calculations are $A_{s} = 0.5$, $t_0 = 1$, $\delta t = 3$; (b) The same at  $w = 1$. }
	\end{figure}

	We see that there are two physically distinguishable cases. The first case of small $w$ is realized when the conditions  $\tau_0 = \frac{2w}{\sqrt{1 - A_{s}^{2}}} \ll \delta t$  and $\tau_1 \sim w \ll \delta t$ are fulfilled, i.e., when the pumping time $\tau_0$ and the time $\tau_1$ of dropping to zero, are small in comparison with the  pulse duration: $\tau_0 \ll \delta t$ and $\tau_1 \ll \delta t$. These conditions mean that  $\sin{\Phi}$ approaches $A_{s} = \sin{\Phi^\star}$ and drops to $0$ for  short periods of time $\tau_0$, $\tau_1$, correspondingly,  in comparison with the pulse duration $\delta t$, thus $\sin{\Phi}$ shows nearly a rectangular form  which coincides with the pulse $I_p(t)$. This case is demonstrated in Fig.\ref{fig:sinPhiIp}(a). Here the magnetic moment feels  an approximately constant field $\sin{\Phi}$ during the pulse.
	
	In the opposite case of $w \gtrsim \tau_0,\tau_1$ the profile of $\sin{\Phi}$ becomes more complicated. First, the pumping process of $\sin{\Phi}$ to $A_{s}$  becomes broader. Second, a significant tail emerges where $I_p(t) = 0$, but $\sin{\Phi} \neq 0$, which influences the magnetization dynamics. This situation is shown in Fig.\ref{fig:sinPhiIp}(b).
	
	So, one can notice that  parameter $w$ measures time of $\Phi$ reaction to the external current. It could be  concluded that for small $w$  in comparison with the characteristic time scale of $I_p(t)$ and for arbitrary current pulse $I_p(t)$, which values are not very close to $1$ (to $I_c$), time derivative $ w \dot{\Phi}$  in (\ref{eq:rsj}) can be neglected and then, the relation $\sin{\Phi} = I_p(t)$ works well.

\begin{figure}[htb!]
		\centering
		\includegraphics[width=4.2cm]{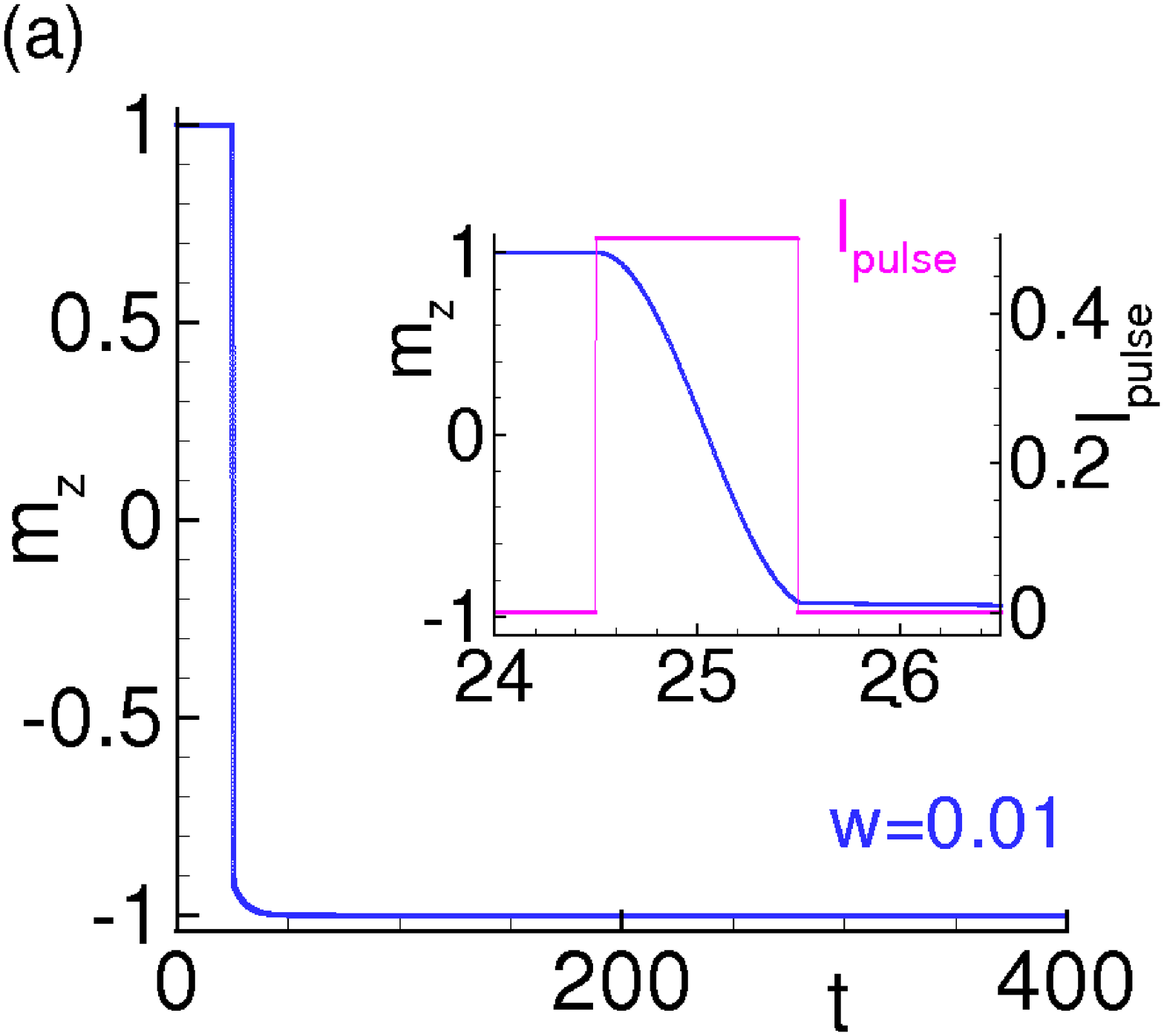}
		\includegraphics[width=4.2cm]{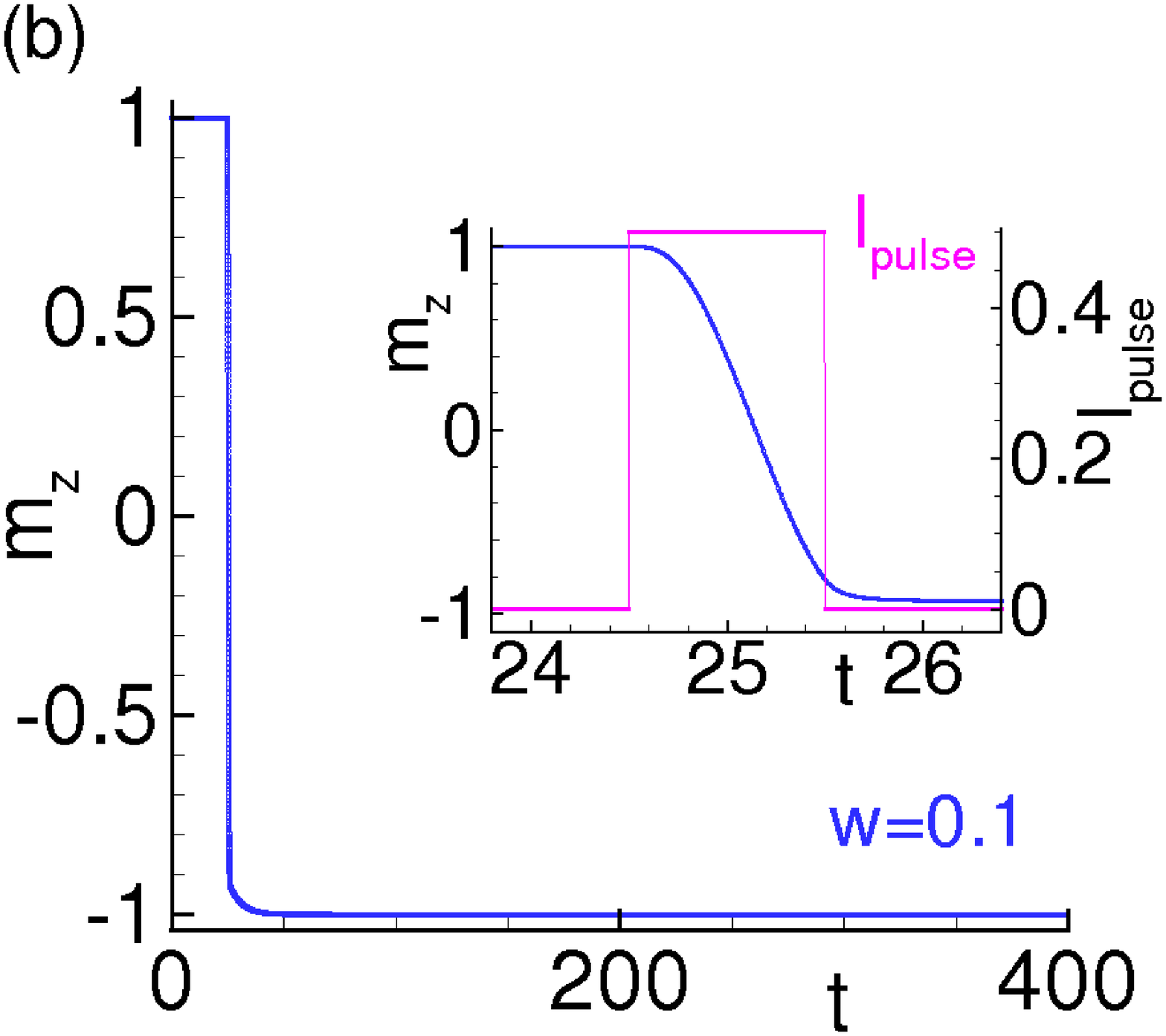}\\
		\includegraphics[width=4.2cm]{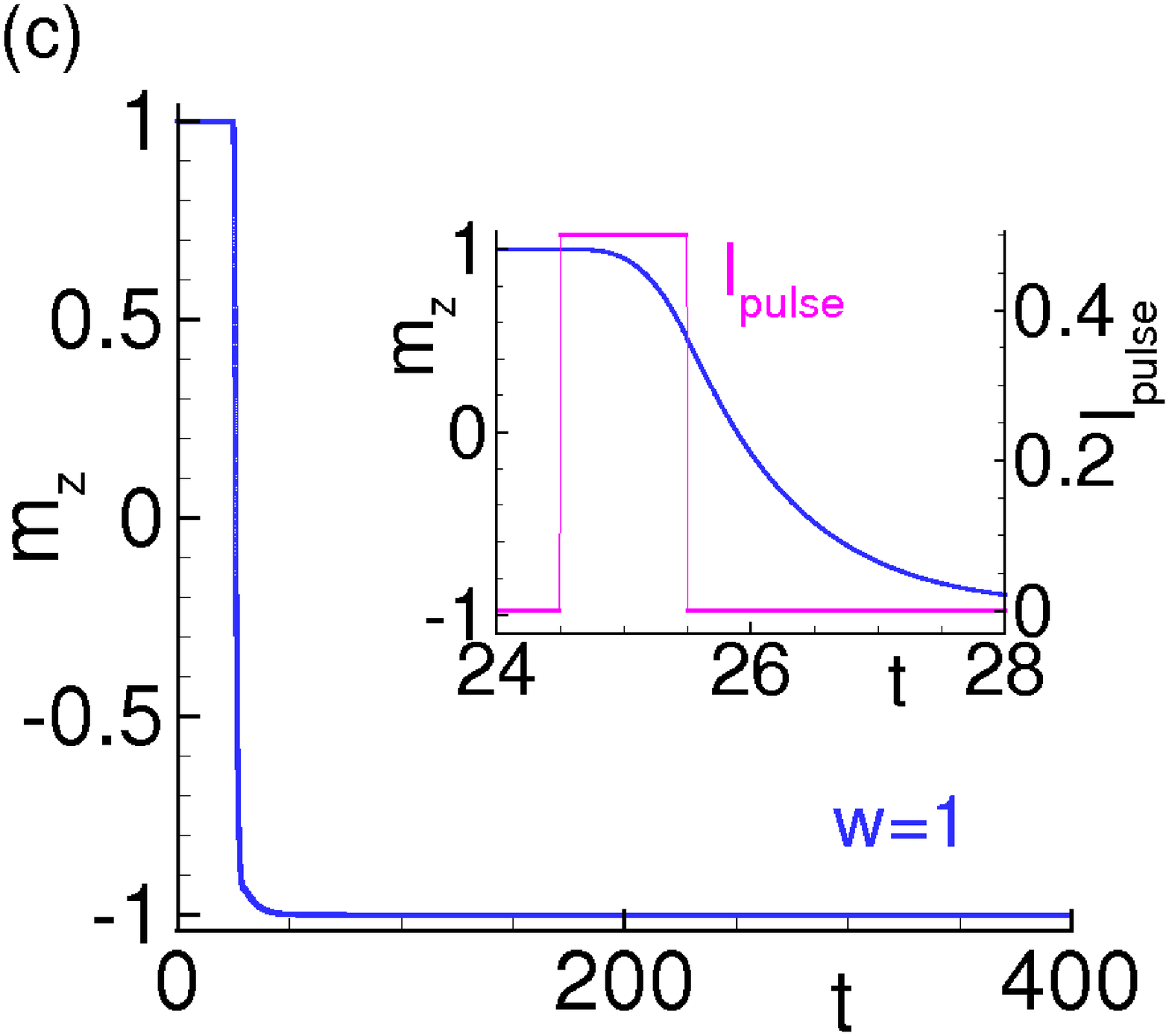}
		\includegraphics[width=4.2cm]{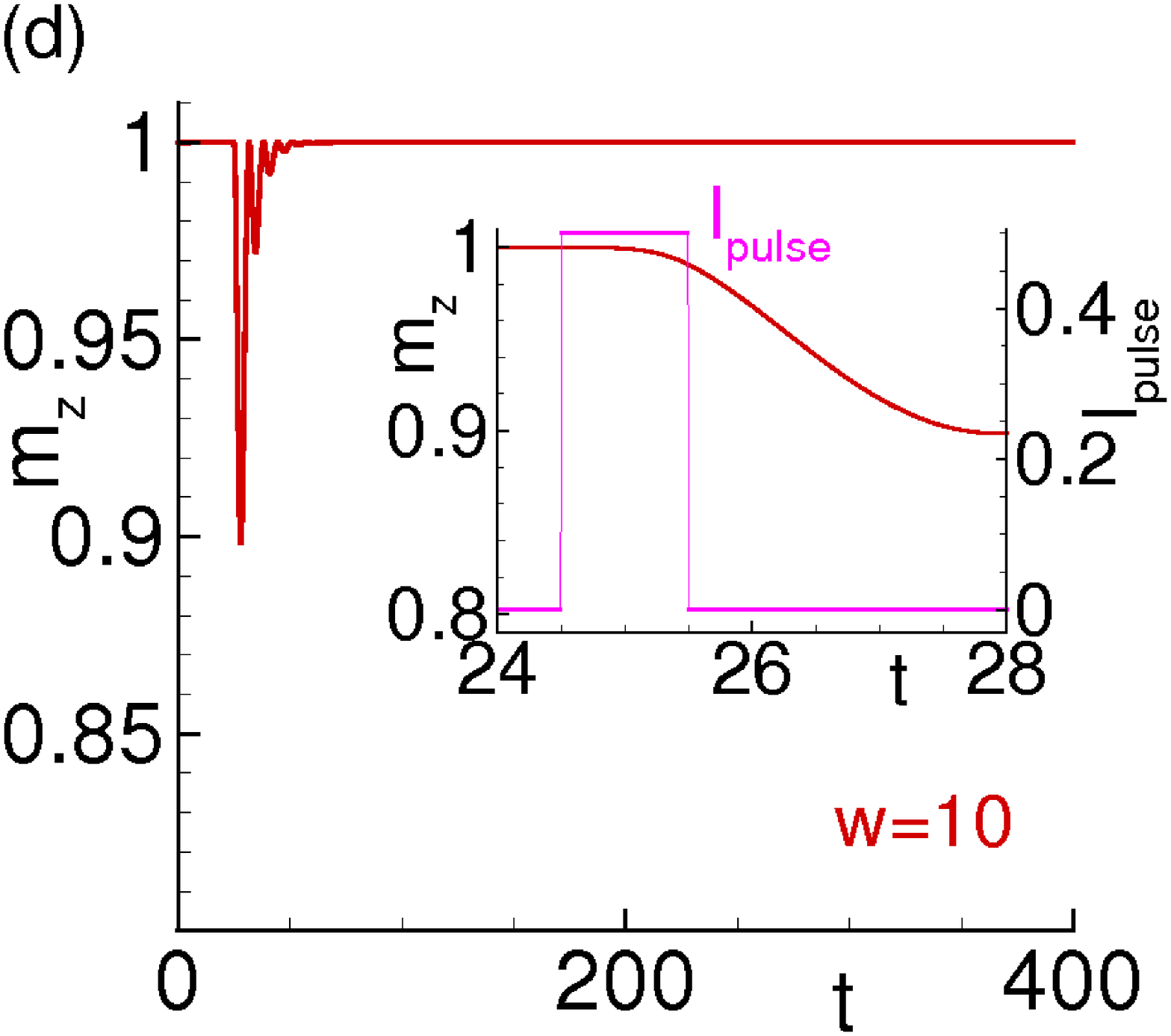}
		\caption{\label{fig:effect_w} Time dependence of $m_{z}$ at $G=60$, $\alpha=0.1$, $r=0.1$, $A_{s} = 0.5$ and $\delta t = 1$  for different values of parameter $w$ indicated in the figures. Insets demonstrate the enlarged parts showing the $m_z$ dynamics in time interval around current pulse.}
	\end{figure}

A question appears: does magnetization reversal is determined by the value of $m_z$ at the end of the current's pulse? The answer is positive for small $w$ only, i.e., when $\omega_F\ll \omega_c$. In this case, when the pulse is switched off, the $y$ component of effective field in LLG equation can be neglected, because $I_p\approx \sin \Phi$. Then, the dynamics of $m_z$ is determined by parameters of LLG equation: if $m_z<0$, we observe the magnetization reversal, and it does not happen in the opposite case ($m_z<0$). At large value of $w$ the magnetization reversal is determined by its tail after switching pulse off. This feature is demonstrated in Fig.\ref{fig:effect_w}, where the influence of $\sin{\Phi}$  tails on  the dynamics of magnetization component $m_{z}$ is shown at different $w$.

In the case $w=0.01$ (see Fig.\ref{fig:effect_w}(a)) and the chosen set of parameters we observe the fastest reversal, i.e. the magnetization reversal happens in the current pulse time interval and after switching the pulse off, the $m_{z} = -1$. In  the case of $w=0.1$, (Fig.\ref{fig:effect_w}(b)) the magnetization reversal is also realized, but the value of $m_{z}$ after the switching  pulse off is $m_{z}=-0.82$ and it reaches the $m_{z}=-1$ at $t=40$. At $w=1$ (Fig.\ref{fig:effect_w}(c)), even the value of $m_z$ is positive and have enough large value at the end of the current pulse ( $m_{z}=0.5$), we nevertheless  observe the magnetization reversal, but it reaches the $m_{z}=-1$ at $t=60$ only. The magnetization reversal is not realized at $w=10$ (Fig.\ref{fig:effect_w}(d)) for the chosen values of the system parameters.  So, the value of $w$, i.e., the relation of $\omega_{F}$ to $\omega_{c}$ plays an important role for the magnetization reversal.

	\section{Solution of LLG equation for $Gr \gg 1$ case}
	Our theory is based on a few key observations. The first one is that for small  $w$ and current pulse $I_p(t)$, which value $A_s$ is not close to $1$ (to $I_c$), as it was discussed in the previous section,  we can neglect the term $w \cdot d \Phi/d t$ in equation (\ref{eq:rsj}),  that implies the relation
	\begin{equation}\label{eq:sinphi}
	I_p(t) = \sin{\Phi}.
	\end{equation}
The second observation is that the condition $w \ll 1$ can be rewritten as $w = \frac{1}{G} \frac{\Phi_0}{2 \pi V} \frac{\gamma \hbar}{2 e R M_0} \sim const \cdot \frac{1}{G} \ll 1$. It means that $w \ll 1 $ does not imply the case of small $G \ll 1$, therefore we can use LLG equation in the $Gr \gg 1$ limit as it was done in Ref.\ \onlinecite{konschelle09}. It was also estimated there, that it is plausible for $G$ to vary in a wide range, starting from $G \ll 1$ till $G \sim 100 \gg 1$.
	
	The third observation is that the Gilbert damping  can be relatively small $\alpha \ll 1$ \cite{weber19,papusoi18,schoen16}. So, if the duration of the current pulse is not long, the damping cannot influence the magnetization significantly, and the system may be considered as it is at $\alpha = 0$. Estimations for this case  is given in the next section.
	
	According to the previous remarks, using (\ref{eq:sinphi}), we can write LLG equation during the pulse as
	\begin{equation}
	\label{syseq1}
	\begin{array}{llll}
	\begin{cases}
	\dot{m}_{x}=Grm_{z}\sin{\Phi } = G r {I_p(t)} m_z,\\
	\dot{m}_{y}=m_{x}m_{z},\\
	\dot{m}_{z}=-Grm_{x}\sin{\Phi} = - G r {I_p(t)} m_x .
	\end{cases}
	\end{array}
	\end{equation}
The limit of the strong coupling $Gr \gg 1$	(but $r\ll1$) can be treated analytically \cite{konschelle09}. In this case $m_y(t) \approx 0$ and for applicability of this method we also need $G r I_p(t) \gg 1$ during the pulse. In the opposite case, zeroes of $I_p(t)$ destroy the predominance of the used terms and more careful consideration should be carried out. Because $m_y(t) \approx 0$, then $m_x = \rho \sin{\phi}$, $m_z = \rho \cos{\phi}$ and  we find directly $\dot{\phi} = G r I_p$. So,
	\begin{equation}
	\begin{array}{ll}
	\phi(t) = G r \int_{t_0}^{t} dt_1  {I_p(t_1)}. \label{eq:mzdin}
	\end{array}
	\end{equation}

As we see from (9), after the pulse has been switched off, the $\sin{\Phi}$ has a fast drop to $0$ due to condition $w \ll 1$. In this time region the dynamics of the magnetization is determined only by the interplay of the magnetic anisotropy and the Gilbert damping, which makes the magnetization to line up along the easy axis \cite{epl18}.

So, as follows from ($\ref{eq:mzdin}$),  the magnetization reversal occurs when
	\begin{equation} \label{eq:cond}
	\cos{\left(  G r \int_{t_0}^{t_0 + \delta t} dt_1  {I_p(t_1)}\right)} < 0,
	\end{equation}
	where $\delta t$ is the pulse duration.
	
\begin{figure}[h!]
		\centering
		\includegraphics[scale=0.3]{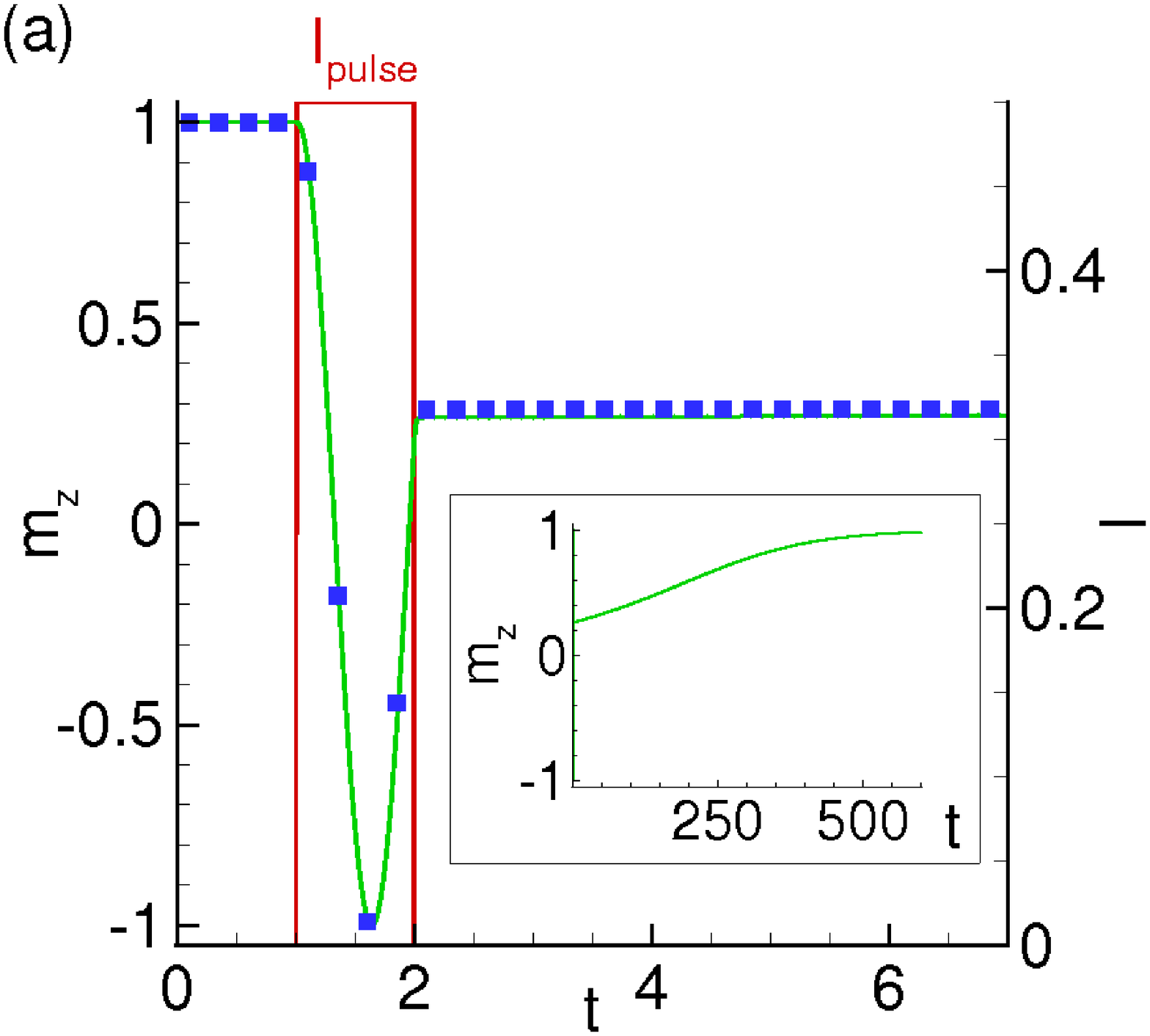}
		\includegraphics[scale=0.3]{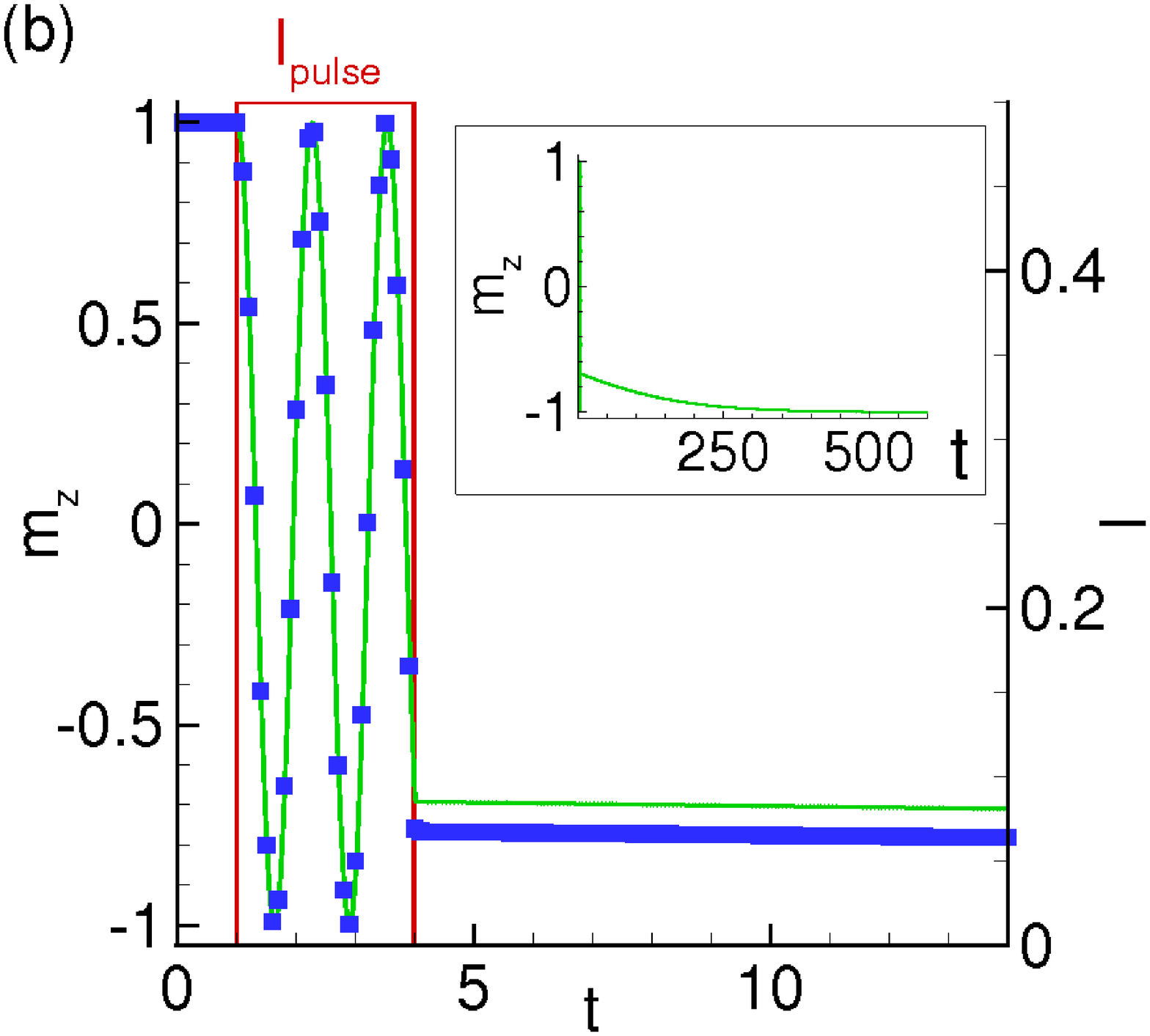}
		\caption{\label{fig:mzdin1}  Dynamics of $m_z$ based on numerical solution of (\ref{eq:LLG}) and (\ref{eq:rsj}) with the effective field (\ref{eq:Heff}) (solid line) and analytical solutions of (\ref{eq:mzdin}) (dashed line) for different pulse amplitude, width and pulse's profile. The current pulse is shown by the red color. The parameters of calculations are $G = 100$, $r = 0.1$, $\alpha = 0.005$, $w = 0.01$, $t_0=1$.
			(a) $A_s = 0.5$ , $\delta t = 1$; (b) $A_s = 0.5$, $\delta t = 3$.}
	\end{figure}

	We illustrate this idea in Fig.\ref{fig:mzdin1}(a) and \ref{fig:mzdin1}(b) for a rectangular pulse $I_p(t) = A_s\left[\theta(t - t_0) - \theta(t - t_0 - \delta t)\right]$ with $A_s = 0.5$ for two pulse durations $\delta t_1 = 1$ and $\delta t_2 = 3$.

\begin{figure}[htb!]
		\centering
		\includegraphics[scale=0.3]{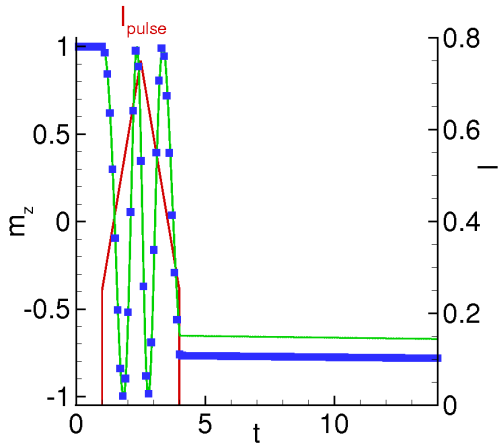}
		\caption{\label{fig:mzdin2}  Dynamics of $m_z$ with the current pulse $I_p(t) = 0.75 - \vert t -t_0 - \delta t/2\vert/3$, $\delta t = 3$. Insets show dynamics after switching current pulse off.}
	\end{figure}	
	
The parameters $G = 100$, $r = 0.1$, $\alpha = 0.005$, $w = 0.01$ are used. In the first case our criteria (\ref{eq:cond}) gives  $\cos\left(G r A_s \delta t_1\right) = 0.28 > 0$, so the reversal is absent, whereas for $\delta t_2 = 3$ we get $\cos\left(G r A_s \delta t_1\right) = -0.76 < 0$ and the reversal occurs.   We see that the solution  (\ref{eq:mzdin}), represented by the blue dashed curve, coincides with the numerical one, represented by the green solid curve, using the complete equations (\ref{eq:rsj}) and (\ref{eq:LLG}) with (\ref{eq:Heff}) during the pulse. When the pulse has been switched off, the damping destroys any deviations from the easy axis $m_z = \pm1$. It is demonstrated in the insets to Fig.\ref{fig:mzdin1}.

It should be noted, that the magnetization reversal is not affected by the form of the current pulse, but by its integral over the pulse duration only. This is demonstrated in Fig.\ref{fig:mzdin2}(c) for the pulse $I_p(t) = 0.75 - \vert t -t_0 -   t/2\vert/3$, $\delta t = 3$. The integral $\int dt_1 I_p (t_1)$ for such pulse is the same as for the pulse in Fig.\ref{fig:mzdin1}(b), so we see that dynamics of $m_z$ and the magnetization  reversal  appearance are not different from the case presented in Fig.\ref{fig:mzdin1}(b).

	\section{Small damping regime }
It was demonstrated in Ref.\cite{jetpl-atanas19} by numerical simulations that there was a periodicity in the appearance of intervals of
the magnetization reversal under the variation of the spin–orbit coupling, Gilbert damping parameter, and Josephson-to-magnetic energy ratio.  Now we can see that the origin of this feature follows from the equation (\ref{eq:cond}) which leads to such periodicity by changing parameters of the system and current pulse.  As a result we will observe the intervals of parameters with the magnetization reversal and its absence. Based on this equation we can reproduce results of numerical simulations of Ref.\cite{jetpl-atanas19} and show the way for optimization of magnetization reversal at different conditions.
	\begin{figure}[htb!]
		\centering	
		\includegraphics[scale=0.3]{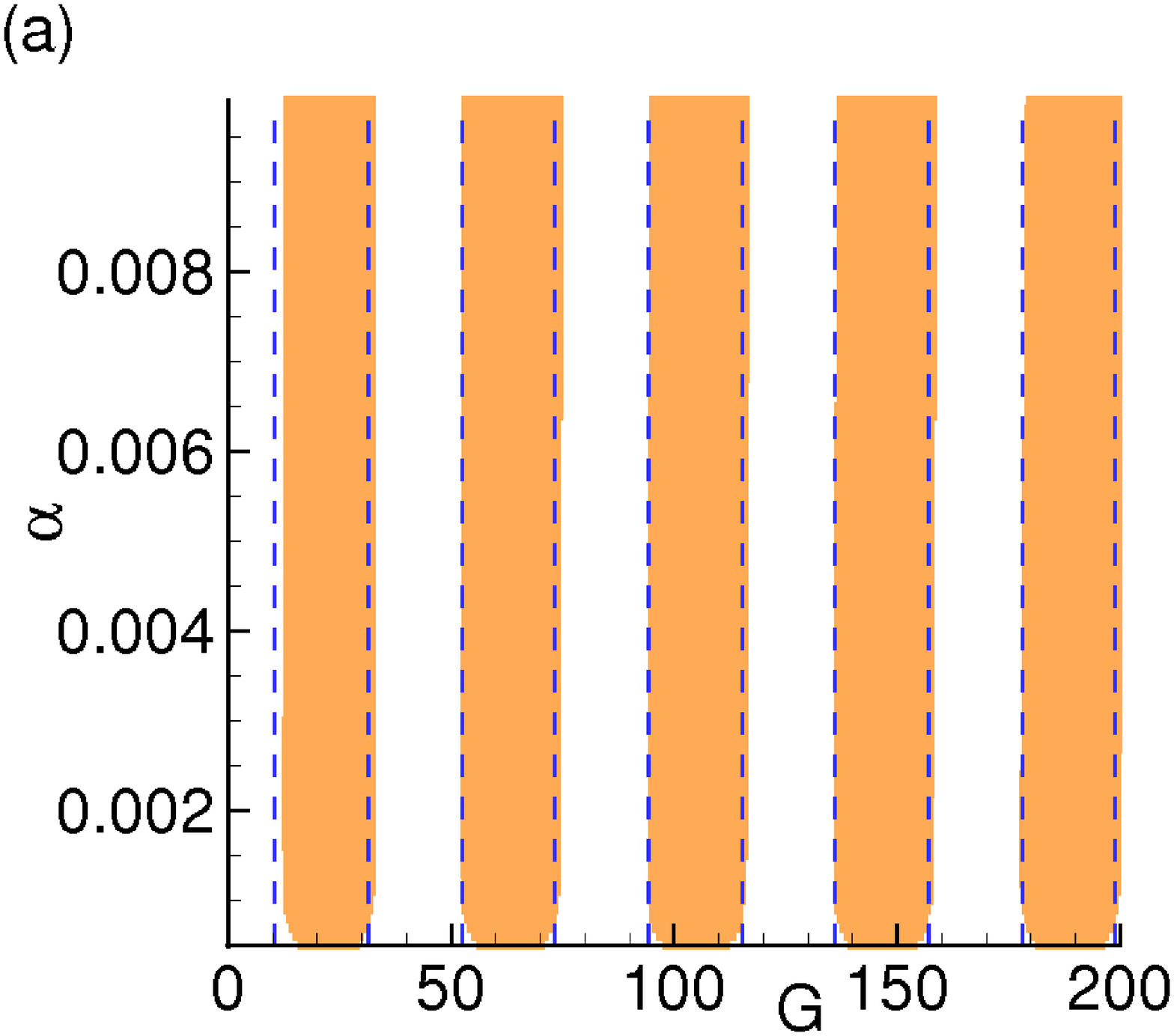}
\includegraphics[scale=0.3]{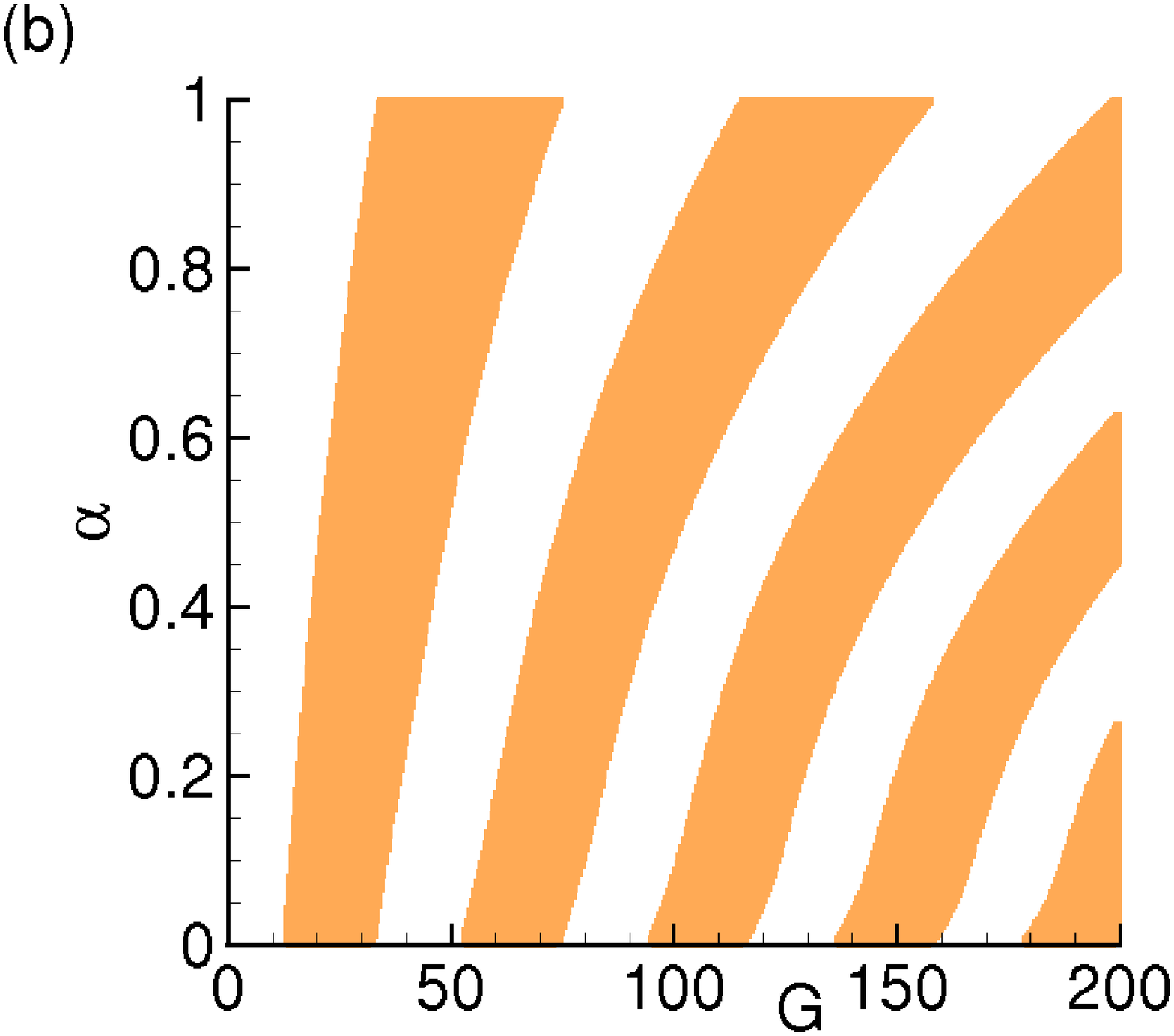}
		\caption{ \label{fig:Galpha} (a) Comparison of analytical and numerical results demonstrating the periodicity MR realization in the $\alpha-G$ plane at small Gilbert damping. The orange stripes reflect the areas  where  magnetization reversal is realized. The dashed blue lines correspond to the areas obtained analytically by (\ref{eq:areas}). The calculation is performed for the $G$  with the step $\Delta G = 0.5$ and for $\alpha$ with the step $\Delta \alpha = 0.0001$. Other parameters are $r=0.1$, $w = 0.01$, $A_{s}/I_c = 0.5$, $\delta t = 3$; (b) Results of numerical calculations at large Gilbert damping produced with the same parameters.}
	\end{figure}
Under current pulse the magnetic moment has complex oscillations determined by system and pulse parameters. Due to the Gilbert damping, the deviated magnetic moment returns back to the stable states with $m_z=1$ or $m_z=-1$. 	To describe its dynamics, we write the equation for $m_y$ including the first non-neglecting term in damping parameter $\alpha$
	\begin{equation}
	\dot{m}_{y}=m_{x}m_{z} + \alpha G r (1 - m^2_y) \sin{\Phi}.
	\end{equation}
	At the beginning of the pulse $m_y(t) \approx 0$, $m_x m_z$ makes fast oscillations due to $Gr \gg 1$, so rising of ${m}_y$ is determined only by the term $\alpha G r \sin{\Phi}$. For applicability of (\ref{eq:mzdin}) we need to keep $m_y(t) \approx 0$, which imposes the condition for the small damping regime
	\begin{equation} \label{eq:alphacond}
	\int_{t_0}^{t_0 + \delta t} dt_1 \alpha G r \sin{\Phi(t_1)} = \alpha G r \int_{t_0}^{t_0+\delta t} dt_1  {I_p(t_1)} \ll 1.
	\end{equation}

For example, at $G=100$, $r=0.1$, $A_s/I_c = 0.5$, $\delta t = 3$ we have $\alpha \ll 0.07$, which corresponds to the experimental value of Gilbert damping parameter \cite{weber19, papusoi18,schoen16}.

According to (\ref{eq:cond}), the magnetization reversal in the $r-G$ plane under pulse $I_p(t) = A_{s}\left[\theta(t - t_0) - \theta(t - t_0 - \delta t)\right]$ occurs in the hyperbolic areas at
	\begin{equation}\label{eq:areas}
	\frac{\pi}{2} + 2 \pi n \leq G_n r {A_{s}\delta t} \leq \frac{3 \pi }{2} + 2 \pi n
	\end{equation}
	for $n = 0, \pm1,...$, whereas the most efficient reversal appears when the condition
	\begin{equation}\label{eq:curves}
	\cos{\left(G r {A_{s}\delta t}\right)} = - 1
	\end{equation}
	is fulfilled, i.e. $G r {A_{s}\delta t} = \pi + 2\pi n$.

Equation (\ref{eq:areas}) does not depend on $\alpha$, but it indicates the intervals of $MR$ at $\alpha=0$. These intervals are shown in Fig.\ref{fig:Galpha} by dashed lines. We see that analytical intervals coincide with the numerical ones, calculated at small $\alpha$. It allows to make a conclusion that magnetization reversal does not depend on $\alpha$ at its small values.

	In order to test this effect of the small damping regime determined by (\ref{eq:alphacond}), we calculate numerically the areas in the $\alpha-G$ diagram where the reversal appears. Results are demonstrated in Fig.\ref{fig:Galpha}(a).

As we see, in the small damping regime the magnetization reversal does not depend on $\alpha$. The areas, where it occurs, are periodic in parameter $G$ , which is determined by(\ref{eq:areas}). As we see, the realization of the magnetization reversal intervals in $\alpha-G$ plane obtained by numerical simulations at small damping, is in agreement with the analytical results.

Results of numerical calculations of the magnetization reversal intervals in $\alpha-G$ plane at large Gilbert damping produced with the same parameters are demonstrated in Fig.\ref{fig:Galpha}(b). We see an essential variations of the stripes at large $G$ and $\alpha$. The estimations made for different SFS Josephson junctions show small value of Gilbert damping when our theory works.\cite{szombati16,aprili19,mayer19} Our theory works also in the limit of small Gilbert damping only.	Nevertheless, we consider that the presented results are the challenge for future theoretical considerations.

	\section{Periodicity of the magnetization reversal in $r-G$  plane}
	
\begin{figure}[h!]
		\centering
		\includegraphics[scale=0.3]{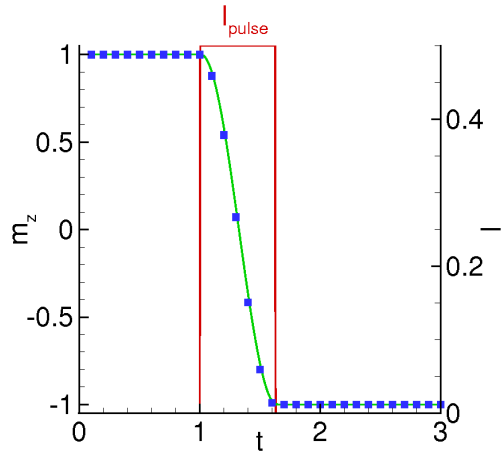}
		\caption{\label{fig:mzdin_fastes}  The same as in Fig.\ref{fig:mzdin1} for the current pulse duration $\delta t_{eff}=0.628$ together with  analytics according to the equations (\ref{t_eff}).}
	\end{figure}

	For simplicity we again consider rectangular pulse at low damping regime and small $w$.
Equation (\ref{eq:areas}) gives the hyperbolic curves for different $n$. From physical point of view they are the curves of a constant amplitude for the driving  force in the LLG equation (\ref{eq:LLG}). In this situation the magnetic moment becomes aligned in the $m_z = -1$ direction exactly after the pulse has been switched off, and the relevant time scale is determined only by the pulse duration, not by the Gilbert damping.  It helps us to optimize the pulse duration in order to make the fastest reversal.  We see from (\ref{eq:curves}) that the shortest time is realized for $n = 0$
	\begin{equation}
	\label{t_eff}
	\delta t_{eff} = \frac{\pi}{G r A_{s}}.
	\end{equation}
	This situation is demonstrated in Fig.\ref{fig:mzdin_fastes} for $G = 100$, $r = 0.1$, $\alpha = 0.005$, $w = 0.01$, $A_{s} = 0.5$ and $\delta t_{eff} = 0.628$. It leads to the reversal time $\delta t_{rev} \approx 0.6\cdot10^{-10}$ s for typical $\omega_{F} \sim 10$ GHz. This time is two orders of magnitude smaller than the estimated one in Ref.\cite{apl17}.
	
	Similar hyperbolic profiles of $1/\delta t_{eff}$ on $I_p(t)$ were obtained theoretically in Ref.\ \onlinecite{sun2000} and experimentally in Refs.\ \onlinecite{nguyen2019,koch2004} for a spin-transfer-induced  magnetization reversal setup in current-perpendicular spin-valve nanomagnetic junctions. In contrast to our case, this type of setup  needs some critical spin-polarized current for magnetization reversal.
\begin{figure}[h!]
		\centering
		\includegraphics[scale=0.25]{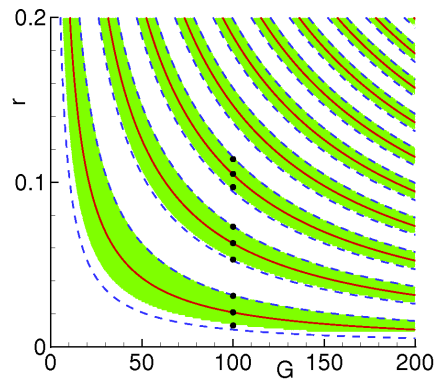}
		\caption{\label{fig:Gr}  Periodicity of the magnetization reversal in the $r-G$ plane. Realization of magnetization reversal is shown by the green stripes, the borders of the areas (\ref{eq:areas}) by the blue dashed lines and the curves for the most efficient reversal (\ref{eq:curves}) by the red lines. The calculation is performed with the step $\Delta G = 0.5$ and step $\Delta r = 0.001$. Other parameters are $\alpha = 0.005$, $w = 0.01$, $A_{s}/I_c = 0.7$, $\delta t = 3$. The solid lines correspond to the analytical expression (\ref{eq:areas}) with $n=0,1,2,...,9$. Black circles indicate the point  where the dynamics of $m_z(t)$ is shown in Fig.\ref{fig:stripe}}
	\end{figure}

\begin{figure*}[htb!]
\centering
\includegraphics[width=4cm]{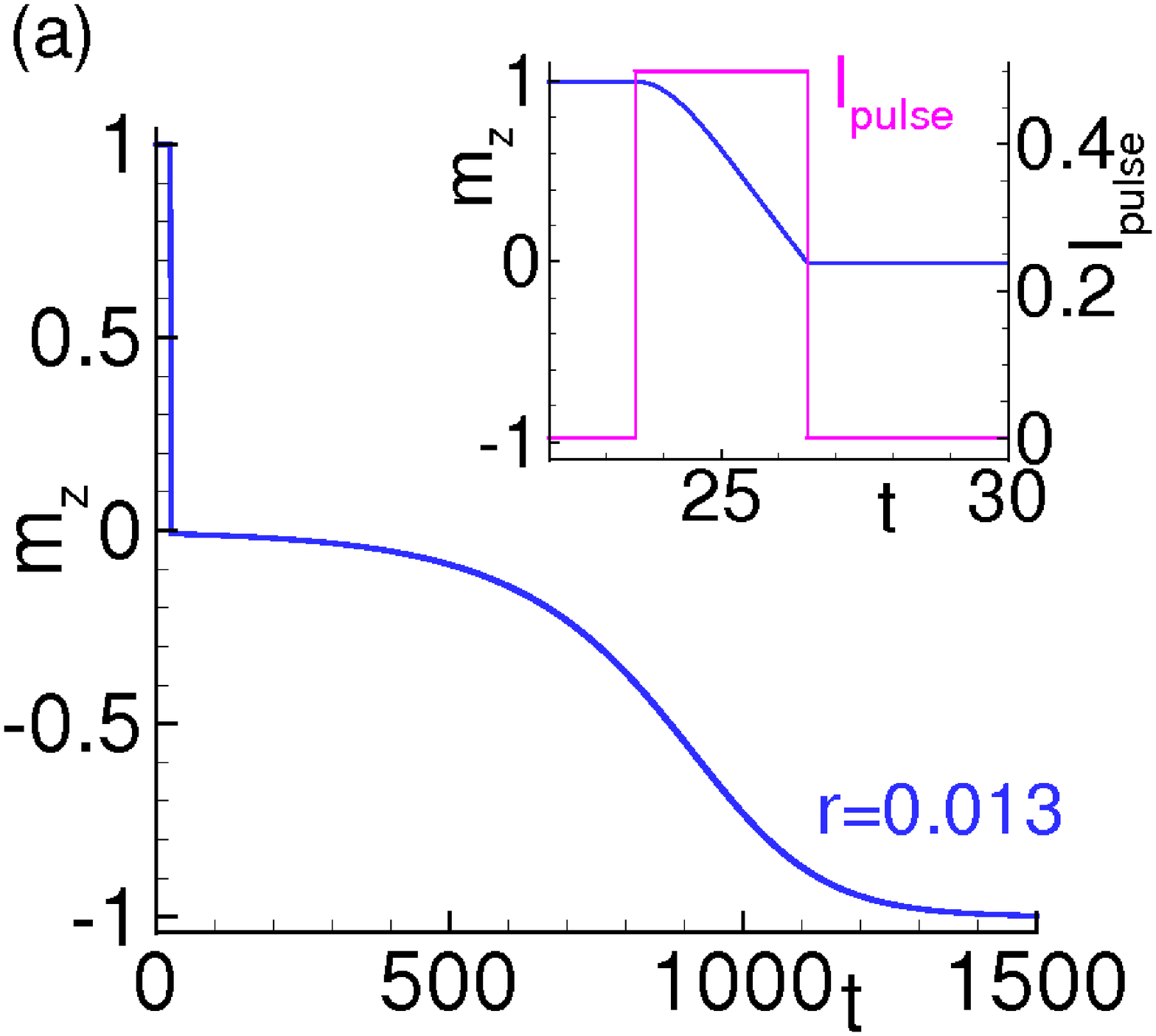}\includegraphics[width=4cm]{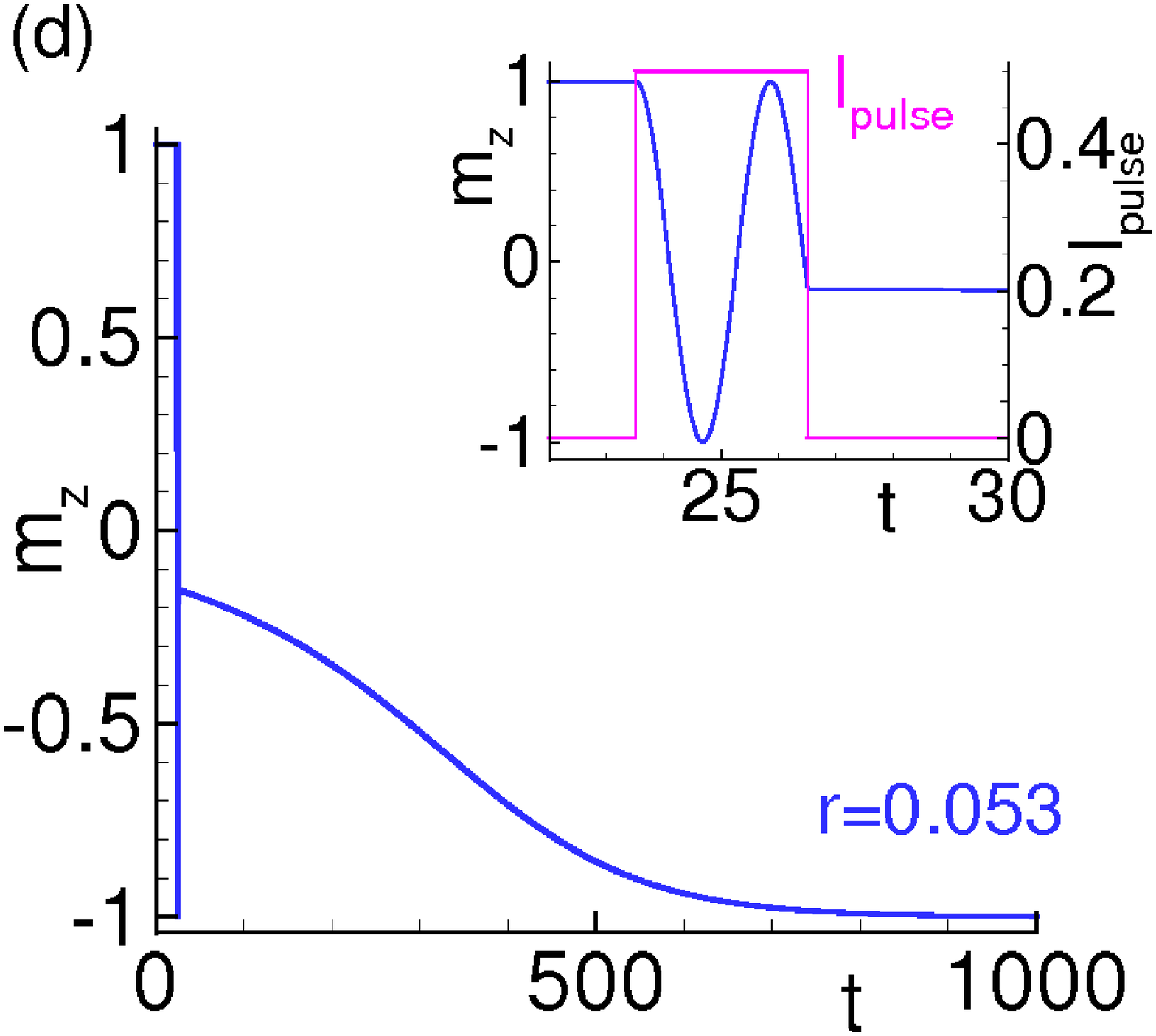}\includegraphics[width=4cm]{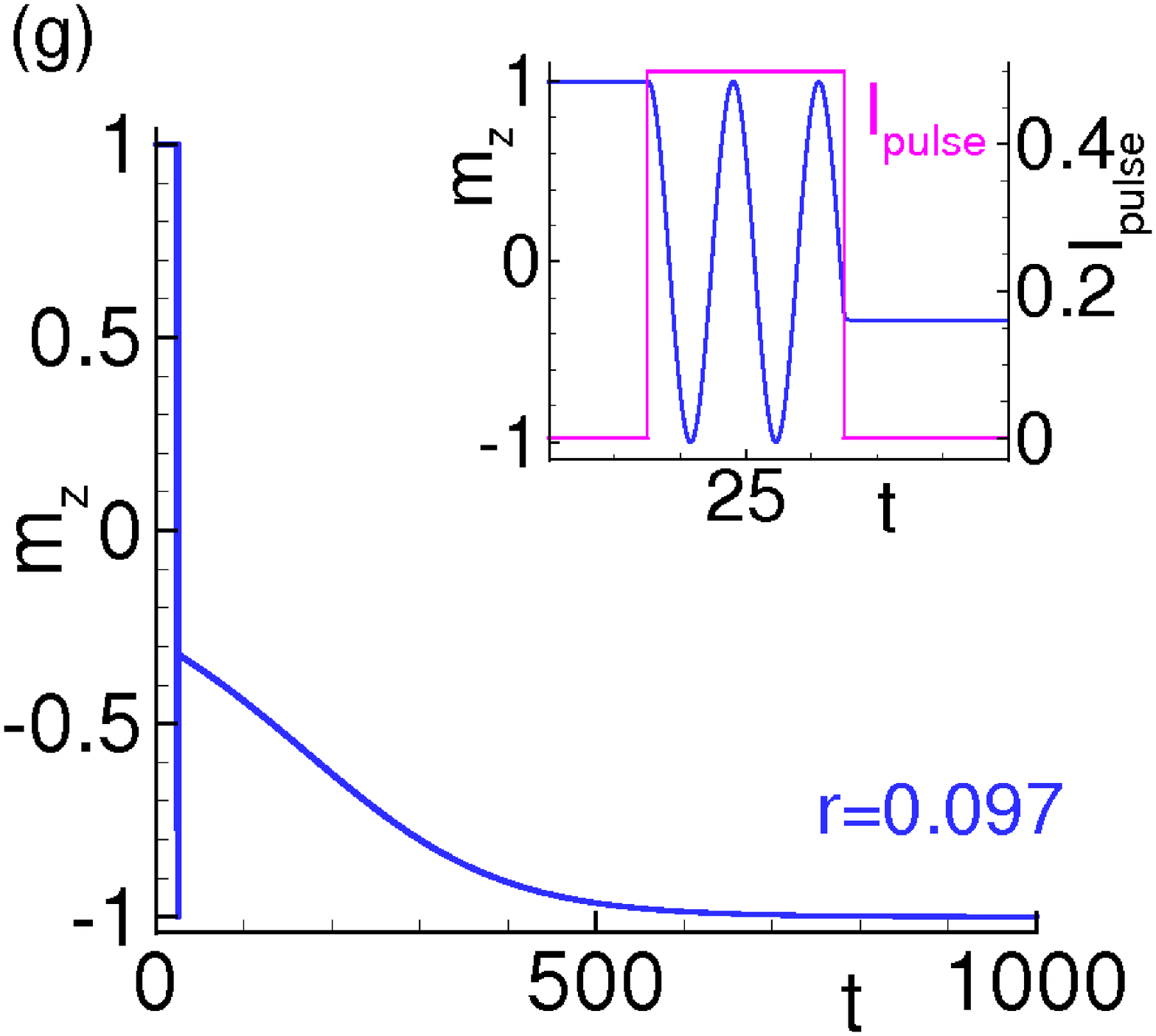} \includegraphics[width=4cm]{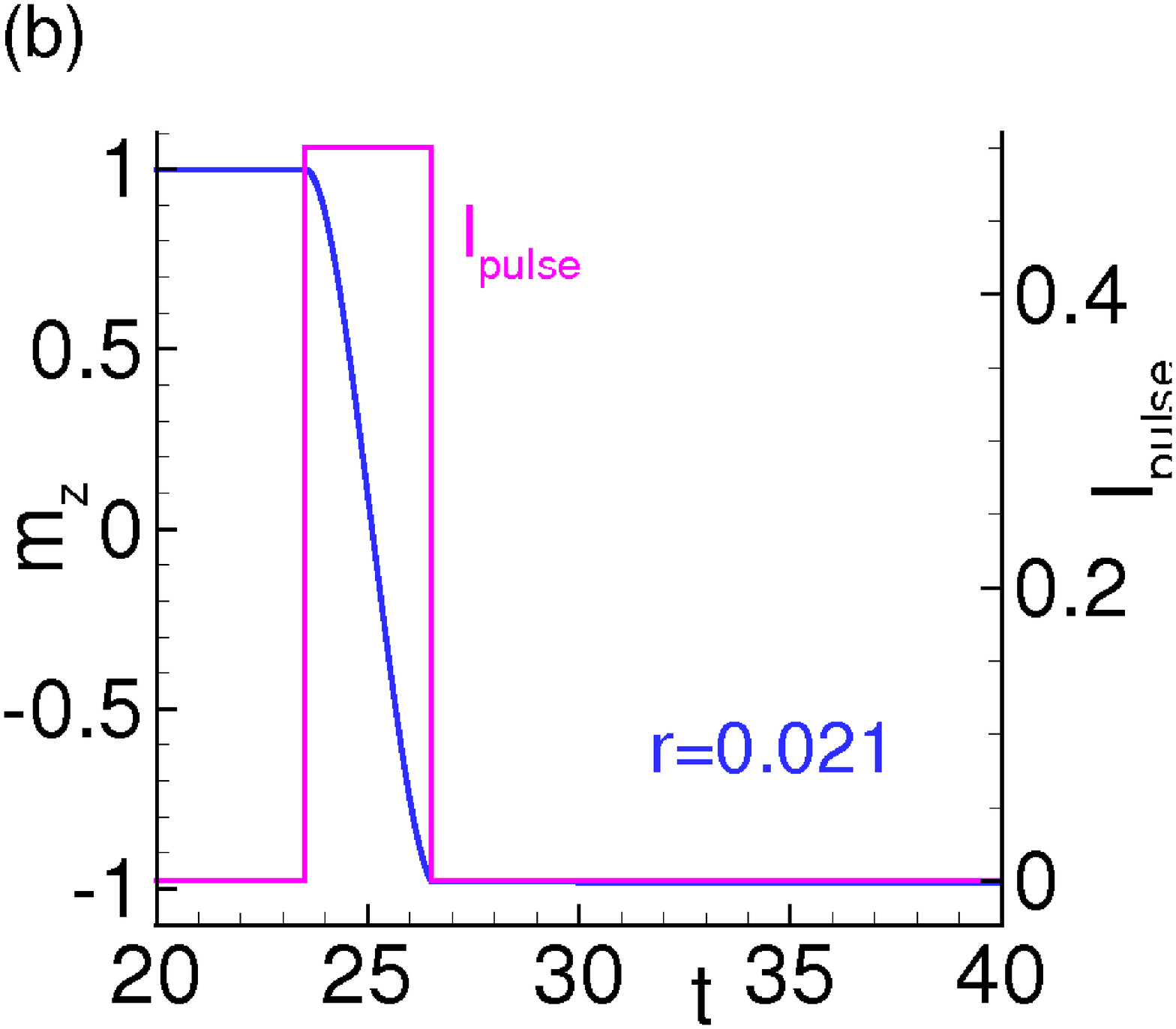}\includegraphics[width=4cm]{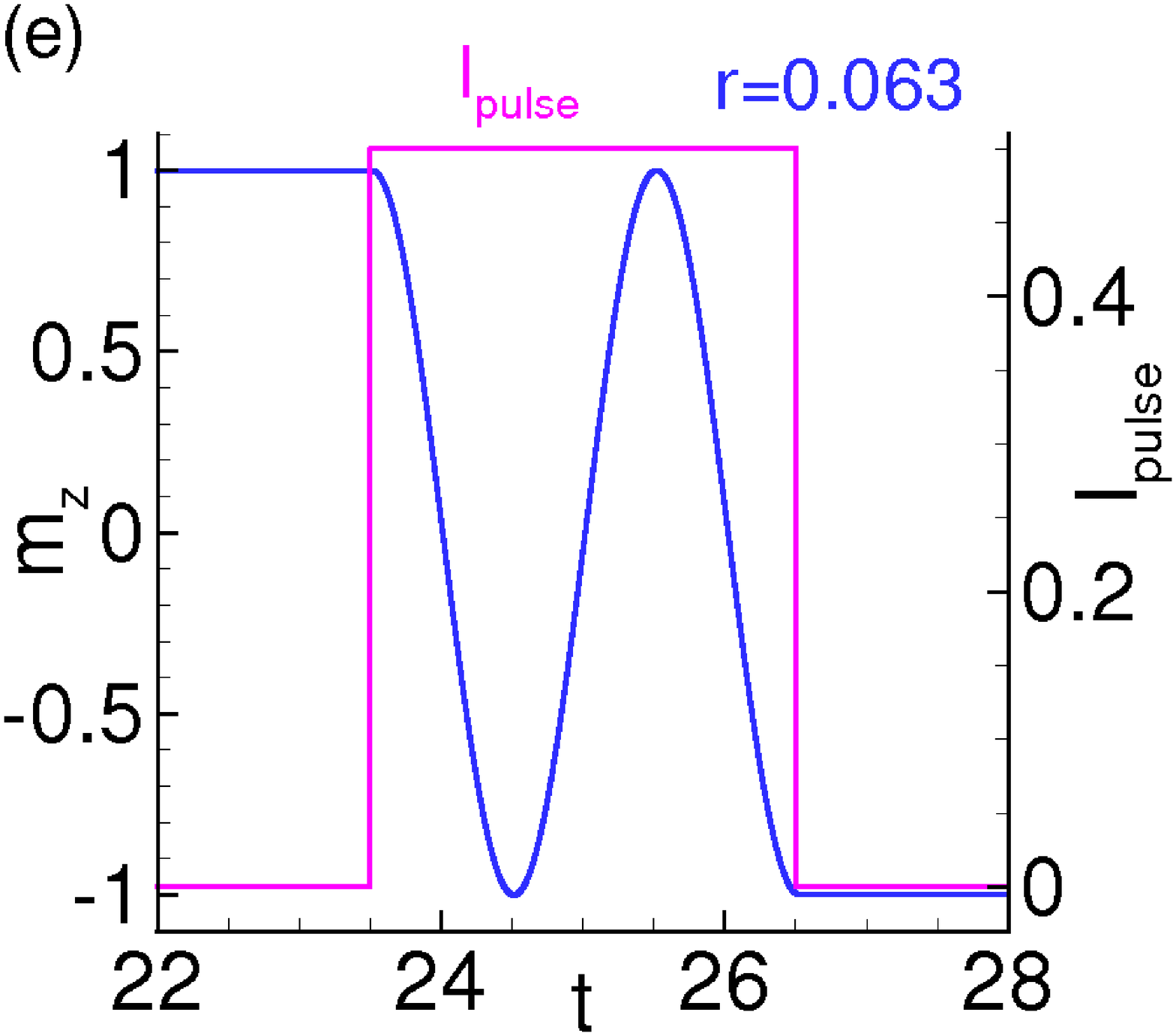}\includegraphics[width=4cm]{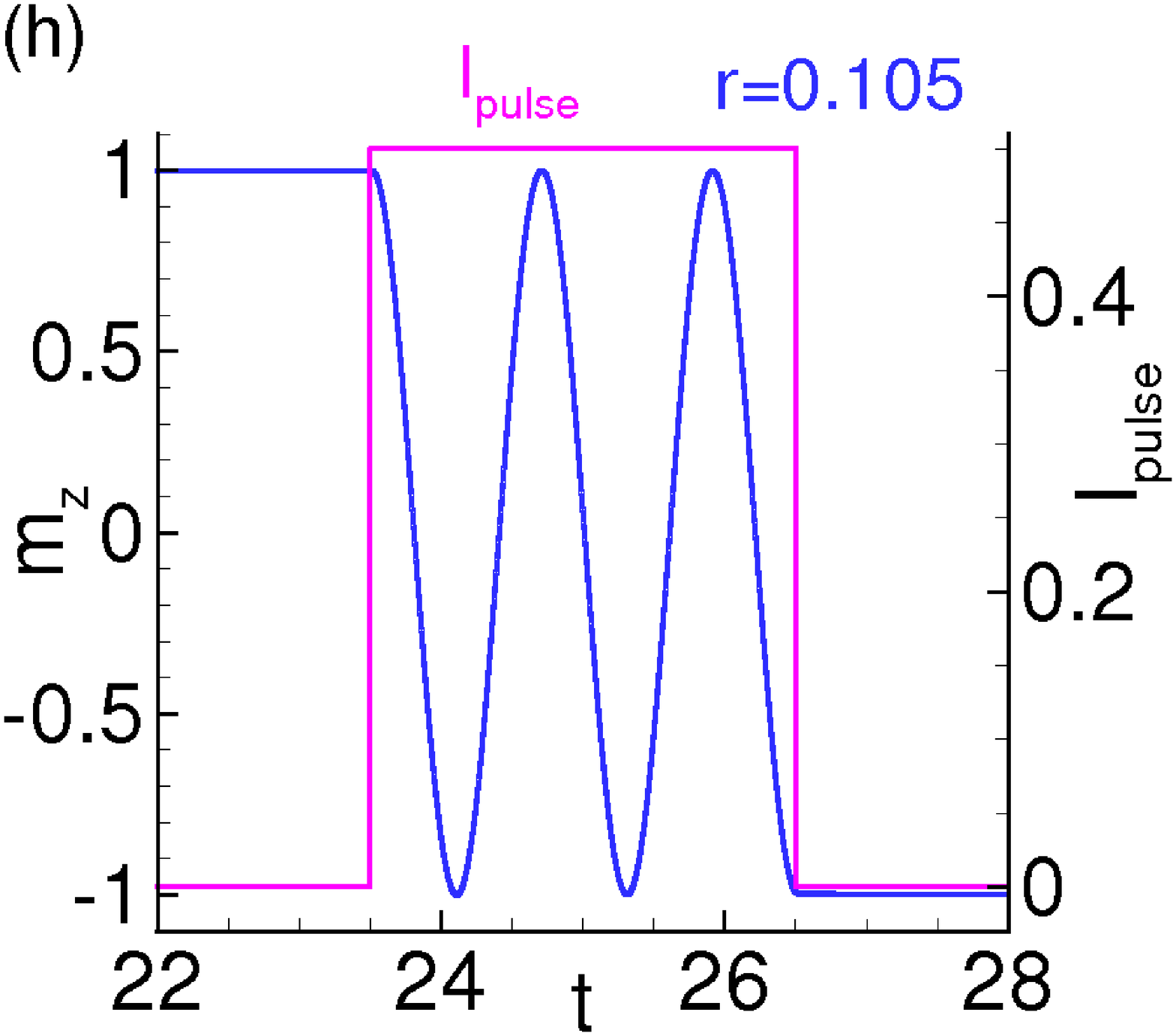} \includegraphics[width=4cm]{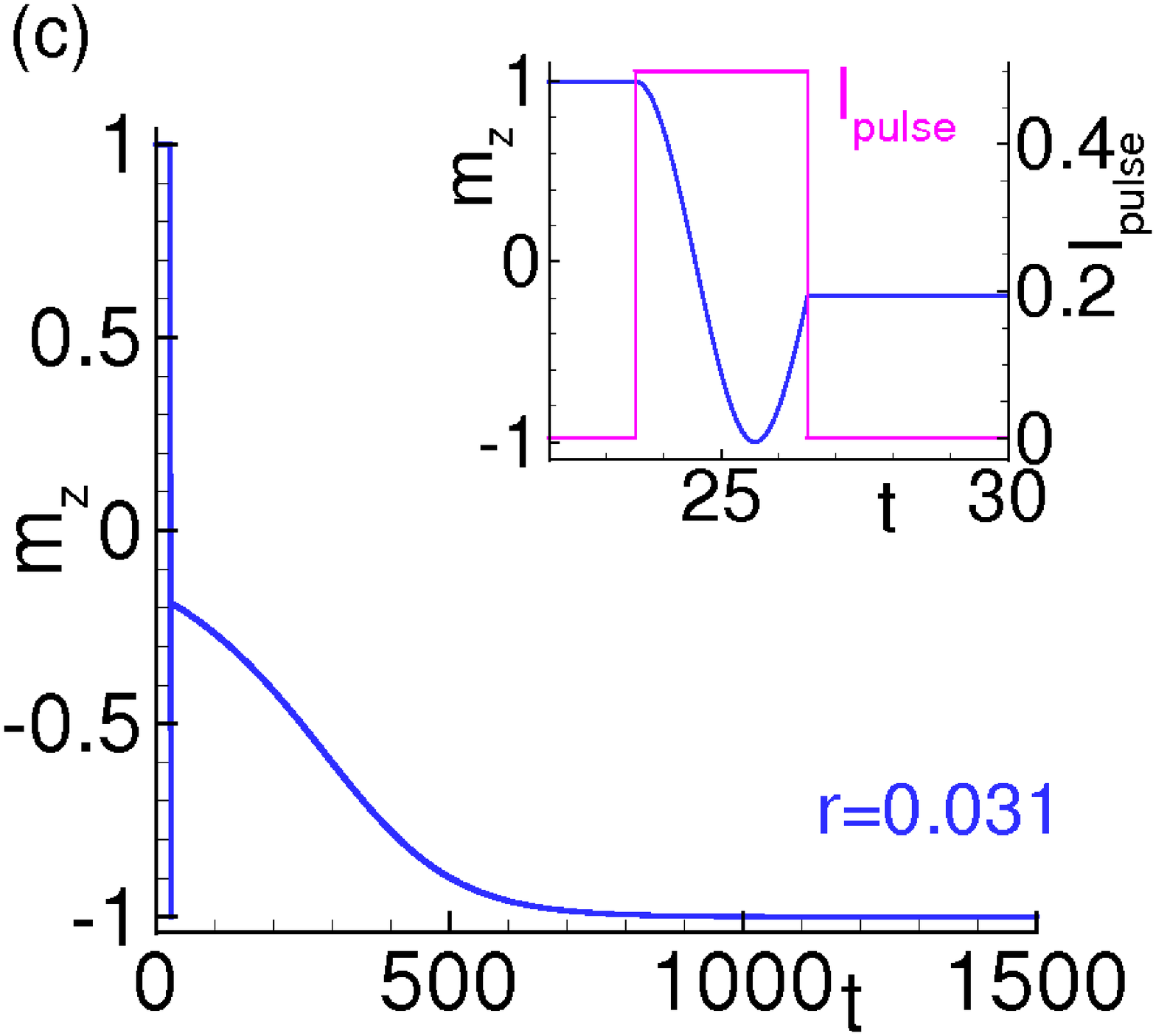}\includegraphics[width=4cm]{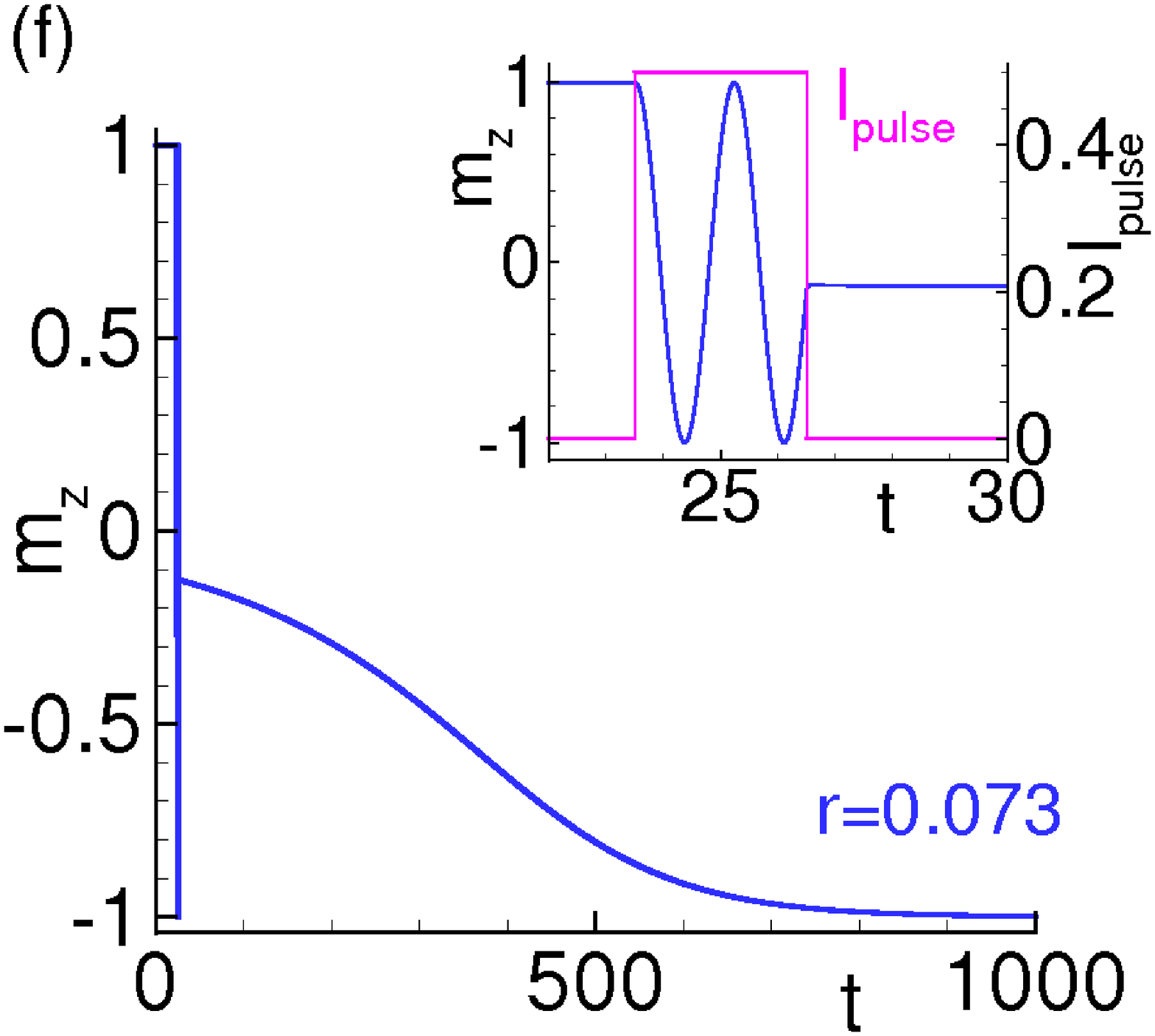}\includegraphics[width=4cm]{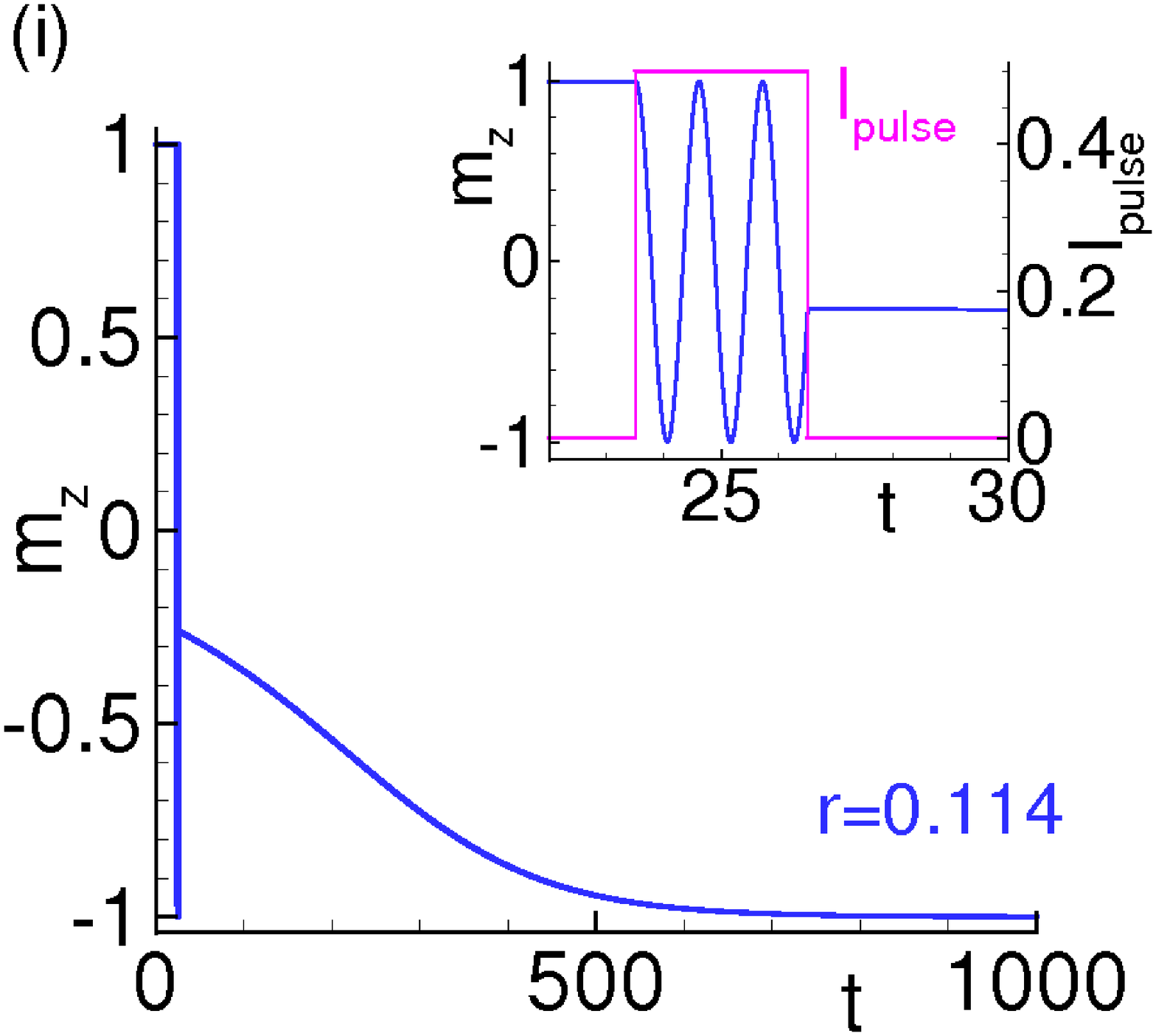}
\caption{\label{fig:stripe} Results of numerical simulations of $m_{z}$ temporal dependence in the first stripe (a,b,c), the second stripe (d,e,f) and  the third stripe (g,h,i) for the set of parameters $G=100$, $\alpha=0.005$ $w = 0.01$, $A_{s} = 0.5$, $\delta t = 3$ and different value of the SO coupling parameters indicated in the figures.}
	\end{figure*}	
	In order to test our analytical results, we calculate numerically the areas of the magnetization  reversal in the $r-G$ plane,  using the complete equations (\ref{eq:LLG}) with (\ref{eq:Heff}) and (\ref{eq:rsj}). In Fig.\ref{fig:Gr} we compare them with the analytical results(shown by dashed lines) based on the equations (\ref{eq:areas}) and (\ref{eq:curves}).

We see the perfect agreement between numerical and analytical calculations which stress the validity of our theory at chosen system's parameters. It should be noticed, that the periodicity of the magnetization reversal in the $r-G$ plane was first observed in Ref. \onlinecite{jetpl-atanas19} numerically only for a non-gauge-invariant scheme. Compare both results we may conclude that, actually, the term $\dot{\varphi}_0$ in (\ref{eq:rsj}), which makes the equations gauge-invariant,  only slightly shifts these areas of magnetization reversal. But, from the other point of view,  the gauge invariant form of equations  gives a possibility for analytical consideration  of equation  (\ref{eq:rsj}).
	
	Finally, we discuss the magnetization reversal at the parameters corresponded to the different points in the stripes, indicated in Fig.\ref{fig:Gr}.  The results of numerical simulations of $m_{z}$ temporal dependence in the first, second and  third stripes at different values of spin-orbit coupling are shown in Fig. \ref{fig:stripe}.	
	
First we compare the magnetization dynamics for three points in the lowest stripe shown in Fig.\ref{fig:stripe}(a,b,c). At the boundaries of this  stripe we observe a slow reversal, while at the point corresponded to the center of stripe ($r=0.021$) the magnetization reversal is the fastest one (Fig.\ref{fig:stripe}(b)). The similar behaviour we observe at the points corresponded to the second and the third stripes shown in Fig.\ref{fig:stripe}. The main important difference between dynamics of the $m_z$ in the centers of the different stripes is following: for the second stripe the $m_z$  makes an additional rotation in compare with a case of the first stripe (see Fig.\ref{fig:stripe}(f)), also for the third stripe $m_z$ makes one more additional rotation  (see Fig.\ref{fig:stripe}(i)). So, it could be directly concluded, that the stripes in Fig. \ref{fig:Gr} differ from each other by number of oscillations made by the $m_z(t)$ component during the current pulse.

It should be noted, that observed periodicity of the magnetization reversal in SFS $\varphi_0$ junction is similar to the well-known effect followed from the Bloch equations in quantum optics and nuclear magnetic resonance\cite{mendel95,jetpl-atanas19}. Generally speaking, we have found here limits of parameters, where the famous $\pi$ pulse is realized in our system. As it could be seen from Figs.\ref{fig:Gr},\ref{fig:stripe} and (\ref{eq:mzdin}), the number of oscillations $n$ made by $m_z$ during the reversal process is proportional to the integral of the pulse function $\int dt I_p(t)$ over time, multiplied by $Gr/\pi$. This property takes place in  our limits for $G r \gg 1$, $w \ll 1$ and small damping regime (\ref{eq:alphacond}), otherwise the process  of the reversal becomes more complicated as it was discussed in Ref.\ \onlinecite{jetpl-atanas19}.

\section{Conclusions}
The $\varphi_0$ Josephson junctions is an interesting and important object for superconducting electronics. Its experimental realization open the way for its different applications, particularly, as a cryogenic memory element. In our paper we have studied the reversal of the magnetic moment in the superconductor-ferromagnet-superconductor $\varphi_0$ Josephson junction and  developed a theory which allows us to understand the phenomena of magnetization reversal  in this system, and also to predict its occurrence at the chosen system's parameters. The analytical criteria for the reversal were derived and tested numerically. We compared analytical results with numerical simulations, explained the observed diagrams $G-r$ and $G-\alpha$ and demonstrated their perfect agreement. We have demonstrated the magnetization reversal at different forms of the current's pulse. In particular, we find the conditions for faster reversal, which is important for the creation of cryogenic
memory based on this system. We consider that the obtained analytical criteria will help
experimentalists to be able to realize memory elements, and will serve as a stimulus
for additional theoretical investigations in this field.

\section{\it Acknowledgment} The reported study was partially funded by the RFBR research projects 18-02-00318
	and 18-52-45011-IND. Numerical calculations were made in the framework of the RSF project 18-71-10095. YuMS and AEB gratefully acknowledge support from the University of South Africa’s visiting researcher program and the SA-JINR Collaborations.


\begin{thebibliography}{99}

\bibitem {linder15}   Jacob Linder  and  W. A. Jason Robinson, Nature Physics  \textbf{11}, 307 (2015).
\bibitem {efetov11}S. Mai, E. Kandelaki, A. F. Volkov, and K. B. Efetov, Phys. Rev.\textbf{B} 84, 144519 (2011).
\bibitem {buzdin05} A. I. Buzdin, Rev. Mod. Phys.  \textbf{77}, 935 2005.
\bibitem {bergeret05}F. S. Bergeret, A. F. Volkov, and K. B. Efetov, Rev. Mod. Phys. \textbf{77}, 1321 2005.
\bibitem {golubov04}A. A. Golubov, M. Y. Kupriyanov, and E. Ilichev, Rev. Mod. Phys.  \textbf{76}, 411 2004.
\bibitem{ghosh17} Roopayan Ghosh, Moitri Maiti, Yury M. Shukrinov, and K. Sengupta, Phys. Rev. \textbf{B 96}, 174517 (2017).
\bibitem {buzdin08} A. Buzdin, Phys. Rev. Lett. \textbf{101}, 107005 (2008).
\bibitem{yokoyama-prb89} Tomohiro~Yokoyama, Mikio~Eto, Yuli~V.~Nazarov, Phys. Rev. B \textbf{89}, 195407 (2014).
\bibitem{minutillo-prb98} M.~Minutillo, D.~Giuliano, P.~Lucignano, A.~Tagliacozzo, and G.~Campagnano, Phys. Rev. B \textbf{98}, 144510 (2018).
\bibitem{krive-prb71} I.~V.~Krive, A. M. Kadigrobov, R. I. Shekhter, and M. Jonson, Phys. Rev. B \textbf{71}, 214516 (2005).
\bibitem{reynoso-prb101} A.~A.~Reynoso, Gonzalo~Usaj,C.~A.~Balseiro, D.~Feinberg, and M.~Avignon, Phys. Rev. Lett. \textbf{101}, 107001 (2008).
\bibitem{alidoust-prb96} Mohammad~Alidoust and Hossein~Hamzehpour, Phys. Rev. B \textbf{96}, 165422 (2017).
\bibitem{Alidoust-prb98-085414} Mohammad Alidoust, Morten Willatzen, and Antti-Pekka Jauho, Phys. Rev. B \textbf{98}, 085414 (2018).
\bibitem{braude-prl98} V.~Braude and Yu.~V.~Nazarov, Phys. Rev. Lett. \textbf{98}, 077003 (2007).
\bibitem{zyuzin-prb93} Alexander~Zyuzin, Mohammad~Alidoust, and Daniel~Loss, Phys. Rev. B \textbf{93}, 214502 (2016).
\bibitem{zyuzin-prb61} A. Zyuzin, B. Spivak, Phys. Rev. B \textbf{61}, 5902 (2000).
\bibitem{Alidoust-prb98-245418} Mohammad Alidoust, Phys. Rev. B \textbf{98}, 245418 (2018).
\bibitem{goldobin-prl107} E.~Goldobin, D.~Koelle, R.~Kleiner, and R.~G.~Mints, Phys. Rev. Lett. \textbf{107}, 227001 (2011).
\bibitem{goldobin-prb91} E.~Goldobin, D.~Koelle, and R.~Kleiner, Phys. Rev. B \textbf{91}, 214511 (2015).
\bibitem{Menditto-prb98} R.~Menditto, M.~Merker, M.~Siegel, D.~Koelle, R.~Kleiner, and E.~Goldobin, Phys. Rev. B \textbf{98}, 024509 (2018).
\bibitem{alidoust-prb87} Mohammad Alidoust and Jacob Linder, Phys. Rev. B \textbf{87}, 060503(R) (2013).
\bibitem{Shapiro-prb98} Dmitriy~S.~Shapiro, Alexander~D.~Mirlin and Alexander~Shnirman, Phys. Rev. B \textbf{98}, 245405 (2018).
\bibitem{spanslatt-prb98} Christian Spanslatt, Phys. Rev. B \textbf{98}, 054508 (2018).

\bibitem {konschelle09}F. Konschelle, A. Buzdin, Phys. Rev. Lett .  \textbf{102} , 017001 (2009)
\bibitem{shukrinov-prb19}Yu. M. Shukrinov, I. R. Rahmonov, and K. Sengupta, Physical Review B 99, 224513 (2019)		
\bibitem{dolcini15} F. Dolcini, M. Houzet, and J. S. Meyer,  Phys. Rev. B92, 035428 (2015).
\bibitem{konshelle15} F. Konschelle, I. V. Tokatly, and F. S. Bergeret,  Phys. Rev. B 92, 125443 (2015).
		






\bibitem{szombati16} D. B. Szombati, S. Nadj-Perge, D. Car, S. R. Plissard, E. P. A. M. Bakkers and L. P. Kouwenhoven, Nature Physics, 12, 568--572 (2016).
\bibitem{aprili19}A. Assouline, C. Feuillet-Palma, N. Bergeal, T. Zhang, A. Mottaghizadeh, A. Zimmers, E. Lhuillier, M. Eddrie, P. Atkinson, M. Aprili, H. Aubin, Nature communications { \bf 10}, 126 (2019).
\bibitem{chudn2016} Eugene M. Chudnovsky, Phys. Rev. B93, 144422 (2016).
\bibitem{chudn2010} L. Cai, E. M. Chudnovsky, Phys. Rev. B. {\bf 82}, 104429 (2010).	


\bibitem{waintal02}X. Waintal and P. W. Brouwer, Phys. Rev. \textbf{B 65}, 054407 (2002).
\bibitem{braude08}V. Braude and Ya. M. Blanter, Phys. Rev. Lett. \textbf{100}, 207001 (2008).
\bibitem{linder83}J. Linder and T. Yokoyama, Phys. Rev. \textbf{B 83}, 012501 (2011).
\bibitem{mayer19}W. Mayer, M. C. Dartiailh, J. Yuan, K. S. Wickramasinghe, E. Rossi, and J. Shabani, Nature Communications, \textbf{11}, 212 (2020)
\bibitem{alicea12}Alicea, J., Reports on Progress in Physics 75,076501,(2012)
\bibitem{fornieri19}Fornieri, A., et. al., Nature, 569, 89 (2019)
\bibitem{ren19} Ren, H., et. al.,Nature, 569, 93 (2019)	
\bibitem{herr11} Q. P. Herr, A. Y. Herr, O. T. Oberg, and A. G. Ioannidis, Journal of applied physics {\bf 109}, 103903 (2011).	
\bibitem{mukhanov11}O. A. Mukhanov, IEEE Transactions on Applied Superconductivity {\bf 21}, 760 (2011).
\bibitem{baek14} B. Baek, W. H. Rippard, S. P. Benz, S. E. Russek and P. D. Dresselhaus, Nature communications {\bf 5}, 3888 (2014).



\bibitem{birge15}N. O. Birge, A. E. Madden and O. Naaman, Spintronics XI {\bf 124}, 10732  (2015).
\bibitem{bergeret19} C. Guarcello and F. S. Bergeret, arXiv:1907.08454 [cond-mat.supr-con] (2019).
\bibitem{nguyen2019} M.-H. Nguyen, G. J. Ribeill, M. Gustafsson, Sh. Shi, S. V. Aradhya, A. P. Wagner, L. Ranzani, L. Zhu, R. Baghdadi, B. Butters, E. Toomey, M. Colangelo, P. A. Truitt, A. Jafari-Salim, D. McAllister, D. Yohannes, S. R. Cheng, R. Lazarus, O. A. Mukhanov, K. K. Berggren, R. A. Buhrman, G. E. Rowlands and T. A. Ohki, Scientific Reports, \textbf{10}, 248 (2020).
\bibitem{apl17}Yu. M. Shukrinov, I. R. Rahmonov, K. Sengupta, and A. Buzdin, Appl. Phys. Lett. \textbf{110}, 182407 (2017).
\bibitem{jetpl-atanas19} P.Kh. Atanasova, S. A. Panayotova, I. R. Rahmonov, Yu.M. Shukrinov, E. V. Zemlyanaya and M. Bashashin, 	JETP Letters \textbf{110}, 722 (2019).
\bibitem{lifshitz91} E. M. Lifshitz and L. P. Pitaevskii, Course of Theoretical	Physics, Theory of the Condensed State Vol. 9 (Butterworth Heinemann, Oxford, 1991).
\bibitem{bobkovi19} D. Rabinovich, I. Bobkova, A. Bobkov, and M. Silaev, 	Phys. Rev. B {\bf 123}, 207001 (2019).		

\bibitem{epl18} Yu. M. Shukrinov, A. Mazanik, I. R. Rahmonov, A. E. Botha, A. Buzdin, Europhys. Lett. { \bf 122}, 37001 (2018).

\bibitem{weber19} R. Weber, D-S. Han, I.Boventer, S. Jaiswal, R. Lebrun, G. Jakob, M. Kläui, Journal of Physics D {\bf 52} 325001 (2019).
\bibitem{papusoi18}C. Papusoi, T. Le, C.C.H. Lo, C. Kaiser and M. Desai, R. Acharya,Journal of Physics D {\bf 51} 325002 (2018).
\bibitem{schoen16} M. A. W. Schoen, D. Thonig, M. L. Schneider, T. J. Silva, H. T. Nembach, O. Eriksson, O. Karis, J. M. Shaw, Nature Physics {\bf 12}, 839 (2016).
\bibitem{sun2000} J. Z. Sun, Phys. Rev. B {\bf 62}, 570 (2000).
\bibitem{koch2004} R. H. Koch, J. A. Katine, and J. Z. Sun, Physical review letters {\bf 92}, 088302 (2004).					
\bibitem{mendel95} L. Mendel and E. Wolf, Optical coherence and quantum optics, Cambridge University Press, Cambridge, UK (1995)

		
	\end{thebibliography}
\end{document}